# From AI Weather Prediction to Infrastructure Resilience: A Real-Time Correction–Downscaling Framework for Tropical Cyclone Impact Forecasting

**You Wu[1], Zhenguo Wang[2], Naiyu Wang[1]**

[1] College of Civil Engineering and Architecture, Zhejiang University, Hangzhou 310058, China

[2] State Grid Zhejiang Electric Power Co., LTD. Electric power Research Institute China

Correspondence: naiyuwang@zju.edu.cn

**ABSTRACT:** This paper addresses a missing capability in infrastructure resilience: turning fast, global AI weather forecasts into asset-scale, actionable risk intelligence. We introduce the AI-based Correction–Downscaling Framework (ACDF), which combines real-time bias correction, terrain-informed downscaling, and fragility-based power transmission system risk assessment for tropical cyclone impacts. ACDF separates storm-scale bias correction from terrain-aware refinement, mitigating error propagation while restoring the sub-kilometer wind variability that governs structural loading. Tested on 11 typhoons affecting Zhejiang, China under leave-one-storm-out evaluation, ACDF produces 500 m wind fields over a province-scale domain, reduces station-scale wind-speed MAE by 38.8% relative to Pangu-Weather, and runs in approximately 25 s per 12-h cycle on a single GPU. In the Typhoon Hagupit case, ACDF reproduced observed high-wind tails, identified a coastal high-risk corridor, and flagged the transmission line that subsequently failed, demonstrating actionable guidance at tower and line scales. ACDF provides an end-to-end pathway from global AI weather forecasts to operational, impact-based early warning for critical infrastructure.

## 1 Introduction

Tropical cyclones (TCs) are among the most destructive natural hazards, repeatedly disrupting power, transport, and other lifeline systems (Esmalian et al., 2022; Roy et al., 2020; Yang et al., 2025). Yet for infrastructure operators, the central challenge is not only forecasting the storm itself, but translating fast-evolving meteorological information into asset-level risk that can support action before failures occur. This paper addresses a missing capability at the heart of civil infrastructure resilience: turning fast, global AI weather forecasts into real-time, sub-kilometer, unbiased wind fields and asset-level failure probabilities that operators can act on. Rather than proposing a new AI forecast architecture, we bridge state-of-the-art AI-based weather prediction (AIWP) with infrastructure risk forecasting, delivering actionable early warning to support resilience planning with operational lead time.

TCs caused more than $1.4 trillion in global economic losses between 1970 and 2019 (representing 38% of all disaster damages; WMO, 2021), with annual losses now exceeding $100 billion and repeatedly disrupting critical infrastructure across coastal regions (China Power, 2019; NOAA, 2019; U.S. DOE, 2018). While AIWP research has been motivated and evaluated largely within the meteorological community—where success is measured in track accuracy, intensity skill, or medium-range lead time (Bauer et al., 2015; Lam et al., 2023; Bi et al., 2023; Chen et al., 2023; Allen et al., 2025; Price et al., 2025)—these products remain poorly aligned with the needs of infrastructure risk management (Reichstein et al., 2025; Huang and Wang, 2025, 2024; Ti et al., 2022; Wu et al., 2025). Mitigating losses in spatially distributed systems such as power grids and transportation networks requires more than forecasts of track or peak intensity. It demands insights into when, where, and with what probability critical assets might fail, and how localized failures propagate into system-level impact—i.e., impact-based forecasting (Coughlan De Perez et al., 2016; WMO, 2021; Harrowsmith et al., 2020; Huang and Wang, 2024; Cui et al., 2025; Yang et al., 2026). This shift from hazard-oriented early warning to infrastructure-aware, impact-based intelligence is increasingly recognized as essential for climate resilience.

Meeting this need requires three capabilities to be satisfied simultaneously in meteorological forecasts: high resolution (sub-kilometer detail to capture terrain-modulated wind variability that governs localized structural loading), scalability (regional coverage to support network-level risk assessments), and computational efficiency (Low-latency inference for real-time decision-making during rapidly evolving TCs).

Physics-based Numerical Weather Prediction (NWP) models struggle to deliver all three. Global systems such as the ECMWF integrated forecasting system (ECMWF-IFS) and the NCEP Global Forecast System (GFS) operate at 9–25 km grid spacing, while even regional systems like High-Resolution Rapid Refresh (HRRR) remain limited to ~3 km—still insufficient for asset-level risk assessment in complex terrain. The Weather Research and Forecasting (WRF) model can achieve <1 km resolution (Skamarock et al., 2019; Perini De Souza et al., 2022), but its computational demands prohibit real-time deployment. Moreover, NWP forecasts often exhibit systematic errors in intensity and storm structure (Rasp et al., 2024; Olivetti and Messori, 2024).

Emerging AIWP models such as GraphCast, Pangu-Weather, Fuxi, Aardvark, and GenCast (Lam et al., 2023; Bi et al., 2023; Chen et al., 2023; Allen et al., 2025; Price et al., 2025) deliver orders-of-magnitude faster inference and competitive medium-range skill. Yet they inherit the coarse resolution and structural biases of their training reanalyses (e.g., ERA5 at ~25 km), limiting their ability to resolve local wind extremes and leading to persistent intensity underestimation (Rasp et al., 2024; Olivetti and Messori, 2024).

This gap is consequential for civil engineering. On one side, AIWP now produces global forecasts in minutes rather than hours, rivaling or surpassing NWP in speed and accuracy. On the other, civil infrastructure systems still lack impact-based risk forecasts—unbiased, high-resolution wind fields coupled to fragility models that predict which assets will fail and when. The result is an implementation gap: fast, powerful AI forecasts exist, but infrastructure communities still lack the means to convert them into actionable risk intelligence for resilience planning.

This study directly bridges that gap. We introduce the AI-based Correction–Downscaling Framework (ACDF), a modular, interpretable system that processes raw AIWP outputs into 500 m, province-scale, near-real-time wind fields and couples them to fragility models to produce asset-level failure probabilities. ACDF decouples storm-scale bias correction from terrain-aware downscaling, mitigating error propagation and aligning forecasts with infrastructure needs. Fig. 1 provides a conceptual overview, and the main contributions of this study are as follows:

- Bridging AI weather prediction and infrastructure risk forecasting. We present, to our knowledge, the first framework that translates state-of-the-art AI forecasts into real-time, infrastructure-specific risk predictions. Unlike prior AIWP work focused on track or intensity, ACDF targets fragility-informed failure probabilities of power transmission systems under TCs.

- A modular correction–downscaling architecture. ACDF decouples storm-scale bias correction from terrain-aware downscaling, improving interpretability, mitigating error propagation, and addressing both large-scale displacement errors and fine-scale terrain effects.

- Province-scale, high-resolution, validated performance. ACDF processes AIWP forecasts to produce 500 m wind field forecasts over domains exceeding 100,000 km² within seconds. In Leave-One-Storm-Out (LOSO) cross-validation on 11 TCs, ACDF achieves a 38.8% MAE reduction vs. Pangu-Weather (selected bench-mark AIWP) forecasts, with consistent gains in high-wind and complex-terrain conditions.

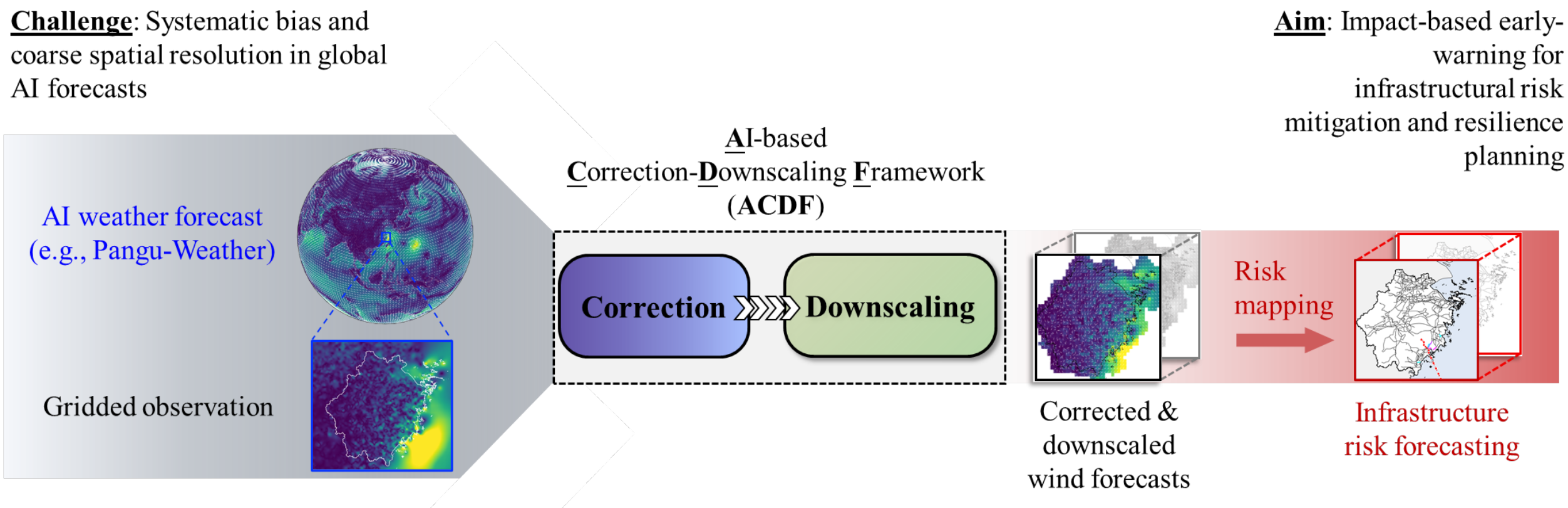

**Fig. 1.** Conceptual overview of the proposed AI-based Correction–Downscaling Framework (ACDF).

**Table 1.** Literature gap summary: wind forecast correction & downscaling.

| Category | Study /Approach | Real-Time Correction | Terrain-Aware Downscaling | Resolution Achieved | Regional Scale (>$10^5$ km²) |
|---|---|---|---|---|---|
| **Site-specific** | Ding et al. (2019); Lim et al. (2022); Wu et al. (2025) | √ | × | Site scale | × |
| **Continuous field** | Huang et al. (2024) | √ | × | 25 km | √ Global |
| | Wu et al. (2024) | √ | × | 3 km | √ Regional, no terrain |
| | Xiang et al. (2024) | √ | × | 3 km | √ CONUS |
| | Liu et al. (2025) | × Post-analysis only | × | 25 km | √ Global archive |
| | Guo et al. (2025) | √ WRF label | × | 25 km | √ Global |
| **Label-free** | Han et al. (2024) | × | √ | 9 km | √ Global |
| | Liu et al. (2025) | × | √ | 3 km | √ Partial CONUS |
| **Label-dependent** | Liu et al. (2024) | × | √ ERA5 label | 25 km | √ |
| | Zhong et al. (2024) | × | √ HRCLDAS label | 1 km | √ Eastern Inner Mongolia, China |
| | Mardani et al. (2025) | × | × WRF label without terrain input | 2 km | √ Taiwan island |
| | Lockwood et al. (2024) | × | × HWIND label | 6 km | × Storm-centric |
| | **This study (ACDF)** | √ **Real-time assimilation** | √ **High-resolution terrain-aware** | **500 m** | √ **Regional-scale (>$10^5$ km², infrastructure networks)** |

- End-to-end infrastructure risk forecast demonstration. In Zhejiang's grid during Typhoon Hagupit (2020), ACDF generated transmission-line-specific failure probability forecasts with lead times up to 12 h that aligned with observed post-event damage and outperformed risk forecasts derived directly from raw AIWP inputs, demonstrating actionable lead time for proactive operations.

ACDF's ability to convert global forecasts into risk intelligence within seconds opens the possibility of delivering impact-based early warnings that save lives and reduce economic losses by enabling targeted evacuations, proactive grid operations, and pre-event resource allocation. To our knowledge, this represents an early demonstration of an end-to-end framework coupling AIWP with risk forecasting for civil infrastructure system at regional scale under tropical cyclones.

## 2 Related Work

Recent studies have improved TC wind forecasts from two main perspectives: forecast correction and wind-field downscaling. Table 1 compares representative correction and downscaling studies from an infrastructure-risk perspective, focusing on real-time applicability, terrain awareness, resolution, and regional scalability.

### 2.1 Forecast correction methods

Forecast correction methods aim to reduce systematic errors in model outputs by incorporating observational information or learning forecast–observation discrepancies. Site-specific methods improve local prediction accuracy using station observations or assimilation strategies (Ding et al., 2019; Lim et al., 2022; Wu et al., 2025), but they cannot directly provide continuous regional wind fields for network-level infrastructure risk assessment. Spatially continuous correction methods further extend correction to gridded fields using GNNs, Transformers, or diffusion-based models (Huang et al., 2024; Wu et al., 2024; Xiang et al., 2024; Liu et al., 2025; Guo et al., 2025). However, most of these methods operate at the native or moderately refined resolution of the input forecasts and generally lack explicit terrain-aware refinement. Thus, although correction can reduce storm-scale bias, it remains insufficient for capturing sub-kilometer wind variability that governs localized structural loading.

### 2.2 Wind-Field Downscaling Methods

Wind-field downscaling methods reconstruct high-resolution fields from coarse atmospheric inputs. Label-free approaches infer fine-scale structures from low-resolution priors or physical constraints without paired high-resolution labels (Han et al., 2024; Liu et al., 2025), but their ability to reproduce terrain-induced local wind variability can be limited. Label-dependent approaches learn coarse-to-fine mappings using paired datasets from reanalysis, WRF simulations or other high-resolution references (Liu et al., 2024; Zhong et al., 2024; Lockwood et al., 2024; Mardani et al., 2025). These methods can recover finer spatial structures, but they are usually designed as offline downscaling models and often assume that the coarse input is already reliable. Consequently, when upstream forecasts contain TC intensity or displacement errors, downscaling may refine spatial details while preserving large-scale bias.

As summarized in Table 1, existing methods address important but separate aspects of the problem. Correction methods improve the large-scale forecast state but usually lack terrain-aware, sub-kilometer refinement. Downscaling methods improve spatial detail but are typically offline and do not assimilate real-time forecast errors. More importantly, most prior studies stop at meteorological field improvement rather than translating corrected and downscaled winds into asset-level infrastructure risk. This gap motivates the proposed ACDF, which decouples real-time storm-scale correction from terrain-aware downscaling and couples the resulting 500 m wind fields with fragility models to support regional, impact-based risk forecasting for transmission infrastructure.

## 3 Formulation of AI-based Correction–Downscaling Framework (ACDF)

From an engineering standpoint, bridging the gap between coarse atmospheric forecasts and infrastructure-scale hazard prediction requires a framework that is both physically interpretable and data-efficient. The AI-based Correction–Downscaling Framework (ACDF) operationalizes this goal through a two-module spatiotemporal learning design that:

1. Assimilates AIWP forecasts and gridded analyses, with high-fidelity station observations used as supervisory labels, so that all available information contributes to forecast refinement;

2. Corrects large-scale, systematic errors in TC intensity and wind structure; and

3. Reconstructs sub-kilometer wind variability driven by static terrain features such as elevation and land cover, which critically govern localized loading on infrastructure assets.

By explicitly decoupling downscaling from correction, ACDF ensures that storm-scale biases are removed before terrain-induced refinements are added, mitigating error propagation and yielding hazard forecasts that are dynamically consistent, geographically realistic, and directly usable for infrastructure risk assessment.

### 3.1 Notation

Table 2 summarizes the primary symbols used in this section. These notations define the inputs, outputs, and intermediate representations of the ACDF.

### 3.2 Problem formulation

ACDF addresses TC wind forecasting using Pangu-Weather (Bi et al., 2023) as the primary source of coarse-resolution inputs. Pangu-Weather is a state-of-the-art AIWP model reported to achieve competitive or superior medium-range forecast skill relative to leading NWP systems while requiring substantially lower inference cost. Given coarse forecasts $\boldsymbol{X}_{\text{pangu}}$, historical gridded observations $\boldsymbol{X}_{\text{LR-obs}}$, station observation data $\boldsymbol{Y}_{\text{ST-obs}}$, and static terrain features $\boldsymbol{X}_{\text{terrain}}$ (notations summarized in Table 2), the framework learns to produce high-resolution forecasts $\widehat{\boldsymbol{Y}}_{\text{ds}}$.

The learning task is formalized as a composite mapping:

$$\widehat{\boldsymbol{Y}}_{\text{ds}} = \mathcal{F}_2\left(\mathcal{F}_1\left(\boldsymbol{X}_{\text{pangu}}, \boldsymbol{X}_{\text{LR-obs}};\ \theta_1\right), \boldsymbol{X}_{\text{terrain}};\ \theta_2\right) \quad (1)$$

Here, $\mathcal{F}_1$ module performs bias correction, generating a coarse corrected field $\widehat{\boldsymbol{X}}_{\text{corr}}$. $\mathcal{F}_2$ module applies terrain-aware downscaling, producing the final 500 m-resolution forecast $\widehat{\boldsymbol{Y}}_{\text{ds}}$.

Physically, $\mathcal{F}_1$ removes TC-scale biases (e.g., intensity misestimation and structure misforecast), while $\mathcal{F}_2$ superimposes localized terrain effects (e.g., ridge speed-up and valley sheltering). This decoupled design enables each module to be independently trained and optimized, ensuring that the final wind field is both dynamically consistent and geographically realistic. Notably, it also allows flexible integration with different upstream forecast models and downstream infrastructure risk assessment workflows.

### 3.3 Correction module ( $\mathcal{F}_1$ ): real-time bias adjustment

The correction module adjusts raw Pangu forecasts so that they remain consistent with the evolving observed atmospheric state. This step is essential because even state-of-the-art AIWP and NWP models can exhibit systematic biases. Here, systematic bias refers to spatially and temporally coherent forecast–observation discrepancies, such as persistent intensity underestimation, displacement of the wind core, or over-smoothed wind gradients, rather than random pointwise noise.

We implement $\mathcal{F}_1$ as a graph-based neural network that processes both the $H$-hour historical segments of $\boldsymbol{X}_{\text{pangu}}$ and $\boldsymbol{X}_{\text{LR-obs}}$ , together with the $\tau$-hour future forecast segment from $\boldsymbol{X}_{\text{pangu}}$ . This sequential encoding captures temporal dependencies and systematic forecast–observation discrepancies between forecasts and observations:

$$\widehat{\boldsymbol{X}}_{\text{corr}} = \mathcal{F}_1(\boldsymbol{X}_{\text{pangu}}, \boldsymbol{X}_{\text{LR-obs}}), \widehat{\boldsymbol{X}}_{\text{corr}} \in \mathbb{R}^{\tau \times N \times 2} \quad (2)$$

The output is a bias-corrected wind field at the original coarse resolution. By aligning large-scale circulation with recent observations, $\widehat{\boldsymbol{X}}_{\text{corr}}$ serves as a dynamically consistent input for terrain-aware refinement in downscaling module.

### 3.4 Downscaling module ($\mathcal{F}_2$): terrain-aware spatial refinement

The second module addresses the inherent scale limitation: even bias-corrected coarse forecasts cannot resolve sub-kilometer wind variability in complex terrain. Features such as steep ridges, narrow valleys, and coastal headlands can locally accelerate or suppress winds, with major implications for infrastructure exposure.

We implement $\mathcal{F}_2$ as a hierarchical encoder-decoder trained on paired coarse wind fields and high-resolution labels generated from offline WRF simulations. Its inputs are the bias-corrected coarse forecast $\widehat{\boldsymbol{X}}_{\text{corr}}$ and high-resolution static terrain maps $\boldsymbol{X}_{\text{terrain}}$:

$$\widehat{\boldsymbol{Y}}_{\text{ds}} = \mathcal{F}_2(\widehat{\boldsymbol{X}}_{\text{corr}}, \boldsymbol{X}_{\text{terrain}}), \widehat{\boldsymbol{Y}}_{\text{ds}} \in \mathbb{R}^{\tau \times H' \times W' \times 2} \quad (3)$$

A residual learning formulation is adopted so that the large-scale flow structures in $\widehat{\boldsymbol{X}}_{\text{corr}}$ are preserved, while $\mathcal{F}_2$ focuses exclusively on reconstructing terrain-induced perturbations. This design ensures that the downscaled forecasts remain physically plausible, geographically realistic, and directly

**Table 2.** Notation for the ACDF framework.

| Symbol | Dimension | Description |
|---|---|---|
| $\boldsymbol{X}_{\text{pangu}}$ | $\mathbb{R}^{(H+\tau)\times N\times 2}$ | Coarse-resolution forecasts from Pangu-Weather, including $H$ hours of history and $\tau$ -hour lead forecasts at $N$ grid nodes; last dimension: zonal ($U$) and meridional ($V$) wind components. |
| $\boldsymbol{X}_{\text{LR-obs}}$ | $\mathbb{R}^{H\times N\times 2}$ | Historical low-resolution wind observations (i.e., HRCLDAS) aligned with the Pangu-Weather grid. (Historical inputs used jointly with forecasts) |
| $\boldsymbol{Y}_{\text{LR-obs}}$ | $\mathbb{R}^{\tau\times N\times 2}$ | Low-resolution grid observations over the forecast period, used as training targets. (Future observations used for supervision) |
| $\boldsymbol{Y}_{\text{ST-obs}}$ | $\mathbb{R}^{\tau\times S\times 2}$ | Station observations at $S$ locations, providing state-based high-fidelity ground truth (used in validation and training supervision). |
| $\boldsymbol{X}_{\text{terrain}}$ | $\mathbb{R}^{H'\times W'\times 2}$ | High-resolution static terrain features (e.g., elevation, land cover) at 500 m resolution, with spatial dimensions $H'$and $W'$. |
| $\widehat{\boldsymbol{X}}_{\text{corr}}$ | $\mathbb{R}^{\tau\times N\times 2}$ | Bias-corrected coarse wind field produced by Stage 1. |
| $\widehat{\boldsymbol{Y}}_{\text{ds}}$ | $\mathbb{R}^{\tau\times H'\times W'\times 2}$ | Bias-corrected and downscaled wind field at 500 m resolution. |
| $\boldsymbol{\mathcal{F}}_1(\cdot;\boldsymbol{\theta}_1)$ | — | Bias correction module, parameterized by $\theta_1$. |
| $\boldsymbol{\mathcal{F}}_2(\cdot;\boldsymbol{\theta}_2)$ | — | Terrain-aware downscaling module, parameterized by $\theta_2$. |
| $\boldsymbol{H}$ | — | Number of historical hours. |
| $\boldsymbol{\tau}$ | — | Forecast lead time (hours). |
| $\boldsymbol{N}$ | — | Number of coarse grid nodes. |
| $\boldsymbol{S}$ | — | Number of weather stations. |
| $\boldsymbol{i}$ | — | Downscaling patch index. |
| $\boldsymbol{n}$ | — | Divided $n$ patches in the d03 region for downscaling. |
| $\boldsymbol{H}', \boldsymbol{W}'$ | — | Spatial dimensions of the high-resolution domain (500 m resolution). |

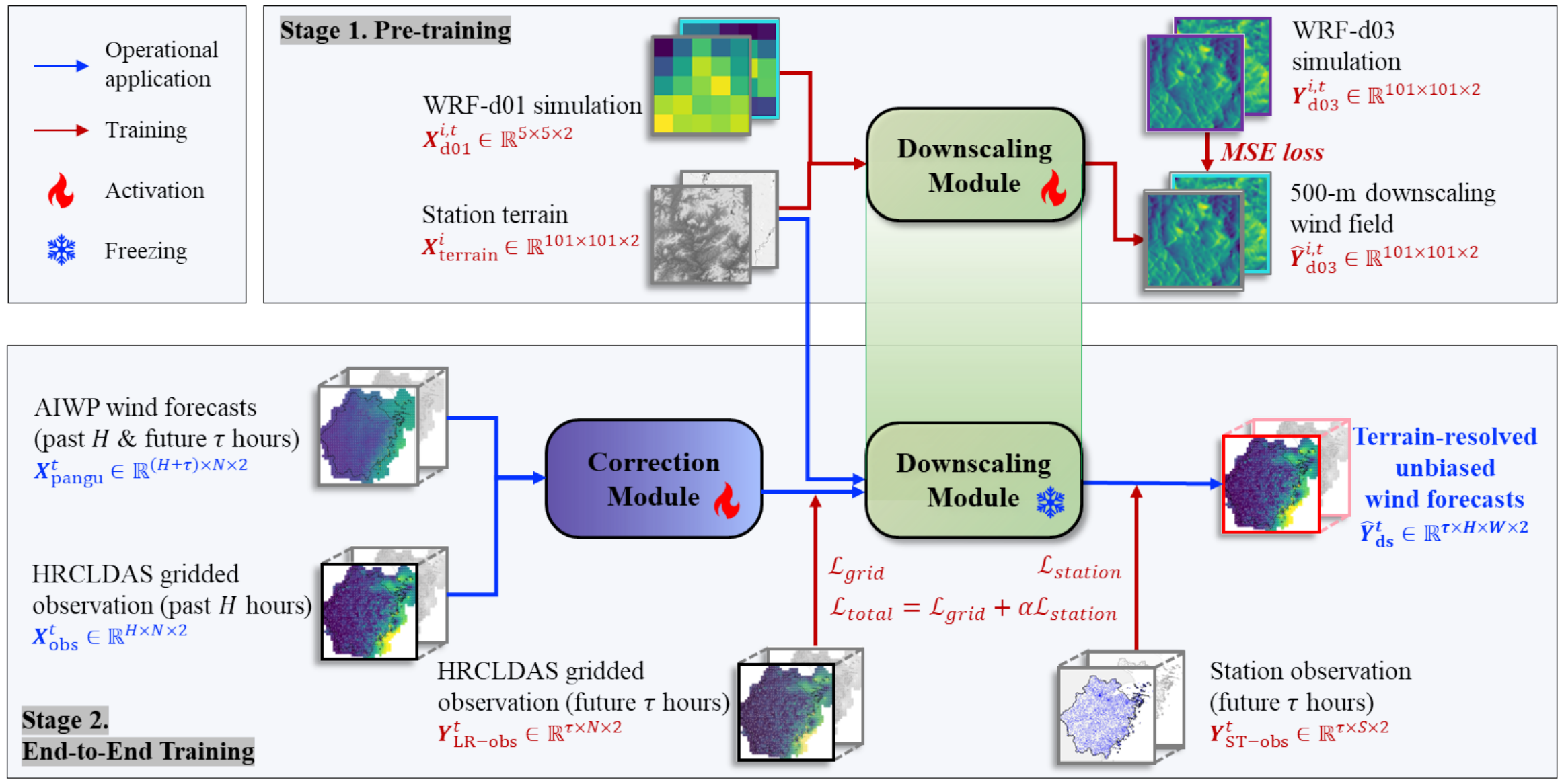

**Fig. 2.** Sequential training strategy of the ACDF.

relevant for infrastructure-scale risk assessment.

### 3.5 Sequential training strategy

To maximize the complementary strengths of the two stages, ACDF adopts a sequential, dependency-aware two-stage training scheme. The overall workflow is illustrated in Figure 2, which highlights how bias correction and terrain-aware downscaling are trained in sequence. This design reflects the distinct physical and statistical roles of each module.

**Stage 1 Pre-train downscaling module ($\mathcal{F}_2$)**

- The downscaling module is trained first using paired coarse-resolution and high-resolution wind fields generated by offline WRF simulations.
- This stage focuses on learning fine-scale, terrain-driven perturbations that are additive to large-scale flow patterns.
- Pre-training in isolation ensures that $\mathcal{F}_2$ develops a stable, domain-general mapping of terrain effects, independent of upstream forecast biases.

**Stage 2 Train correction module ($\mathcal{F}_1$) with parameter-frozen downscaling**

- Once $\mathcal{F}_2$ is trained, its parameters $\theta_2$ are frozen.
- The bias correction module $\mathcal{F}_1$ is then trained with $\mathcal{F}_2$ appended, so that coarse-scale bias adjustments are directly optimized for their impact on the final high-resolution forecasts $\hat{\boldsymbol{Y}}_{\mathrm{ds}}$. This transfer setting is evaluated explicitly in Section 6.4, where we compare input-domain distributions, Stage-2 freezing strategies, and deployment-input performance.
- This dependency-aware strategy prevents systematic errors in storm-scale forecasts from propagating into terrain-refined outputs, as shown in Figure 2.This sequential design operationalizes the core novelty introduced in Section 1: decoupling storm-scale bias removal from terrain-driven downscaling. In this setting, error propagation occurs when intensity or displacement errors in the coarse AIWP forecast are directly passed into the downscaling stage and then reflected as misleading local wind variations. ACDF mitigates this issue by first training the downscaling module to learn terrain-induced perturbations from WRF coarse–fine pairs, and then training the correction module with the downscaling module appended and frozen. As a result, storm-scale biases are corrected before terrain-driven refinements are added, yielding forecasts that remain operationally robust, physically interpretable, and directly applicable to infrastructure risk assessment.

## 4 Data and Preprocessing

Bridging the gap between coarse atmospheric forecasts and infrastructure-scale hazard prediction requires datasets that reflect both storm-scale dynamics and fine-scale terrain effects. To meet this need, we assemble a multi-source dataset that integrates AIWP forecasts, mesoscale numerical simulations, gridded analyses, ground-based observations and static terrain features. Together, these heterogeneous sources serve two complementary roles: (i) enabling the correction module to reduce systematic forecast biases by anchoring predictions to recent observations, and (ii) allowing the downscaling module to reconstruct physically plausible sub-kilometer wind variability shaped by topography. The study area is Zhejiang Province, China— a coastal region with complex terrain and frequent TC exposure, making it an ideal testbed.

### 4.1 Data sources and event set

The multi-source dataset is designed to cover complementary aspects of TC wind forecasting: AIWP provides fast operational forecasts, WRF supplies physically consistent high-resolution supervision, HRCLDAS provides observation-assimilated gridded analyses, station records offer independent point-scale measurements, and static terrain data describe the local controls on wind variability.

This study leverages 11 TCs that impacted Zhejiang Province between 2012 and 2022, covering diverse range of tracks and intensities (Figure 3). These events align with ACDF's two-module design: Five severe TCs with high-fidelity WRF simulations form Group A, supplying paired coarse–fine training data for the downscaling module. All 11 events, including both Groups A and B, contribute forecasts and observations for training the correction module (Table 3). This dual structure ensures both broad event coverage and the high-resolution supervision needed to capture terrain-driven variability.

**AIWP forecasts (Pangu-Weather)**: Hourly 10-m wind forecasts from Pangu-Weather (Bi et al., 2023) provide the coarse, biased inputs that ACDF is designed to refine. The native 0.25° forecasts are bilinearly resampled to a 0.125° (~12.5 km) grid with $N$ nodes, forming tensors $\boldsymbol{X}^t_{\text{pangu}} \in \mathbb{R}^{N\times 2}$ that contain zonal ($U$) and meridional ($V$) wind components. To remain consistent with Pangu-Weather's training baseline, ERA5-driven reforecasts are used in this study, while an operational setting would substitute real-time GFS-driven forecasts.

**High-resolution WRF simulations**: Training data for terrain-aware downscaling are supplied by nested WRF simulations of the five Group A TCs. As a state-of-the-art model for high-resolution weather simulations and used operationally by several national weather agencies, WRF provides physically consistent wind fields that can explicitly resolve terrain effects when configured at km-scale resolution. In this study, three domains were configured: d01 at 12.5 km, d02 at 2.5 km, and d03 at 500 m over Zhejiang Province (Figure 4a). Samples from d01 ($\boldsymbol{X}^t_{\text{d01}}$, 12.5 km) and d03 ($\boldsymbol{X}^t_{\text{d03}}$, 500 m) provide physically consistent coarse–fine pairs, where d01 provides the baseline coarse fields (Figure 4c) and d03 provides terrain-resolved labels (Figure 4e) of how topography modulates TC wind structure. Compared with observation-interpolated wind fields from sparse meteorological stations, the WRF simulations (Figures 4(c-e)) provide dense, coherent supervision across the full domain.

**HRCLDAS gridded analyses**: The HRCLDAS product (Han et al., 2020) provides hourly 1-km wind analyses across mainland China, making it one of the highest-resolution hourly gridded analysis datasets currently available for this region. HRCLDAS is used as an observation-assimilated analysis and training reference, but it is not a substitute for Pangu-Weather as a forecast driver because future HRCLDAS fields are unavailable at operational forecast time. For ACDF, these fields are aggregated to 12.5 km (as in Figure 4g) to align with Pangu's processed grid. HRCLDAS

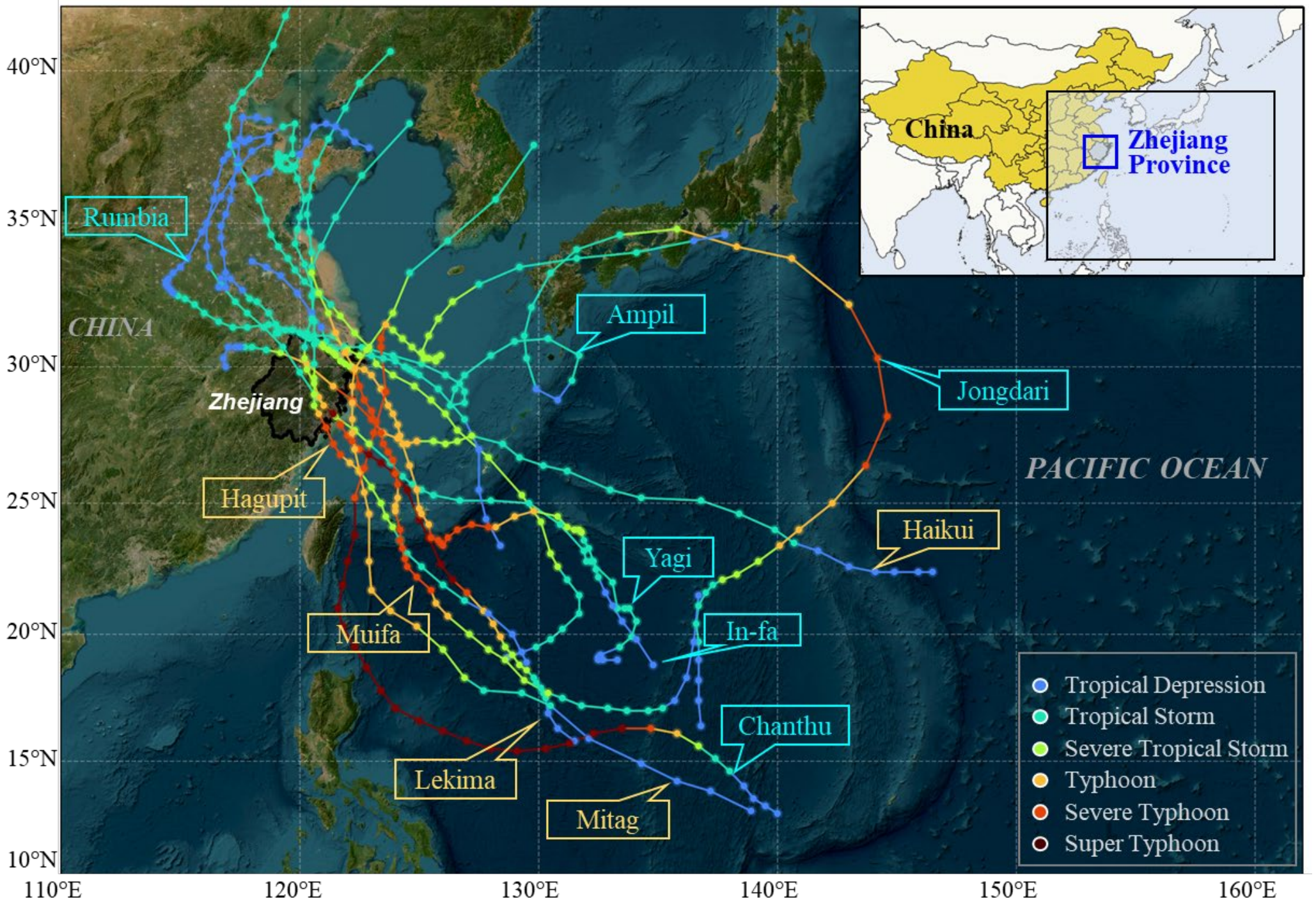

**Fig. 3.** Total 11 TC events used for model development and validation and 5 severe events (in yellow) selected for WRF simulations.

**Table 3.** TC datasets for model development.

| ID | Name | Data Type | Time Span (UTC) | Spatial Coverage* | Notation |
|---|---|---|---|---|---|
| **Group A:** Events with WRF Simulation + Forecast & Observation | | | | **Used in Correction & Downscaling Modules** | |
| 1211 | Haikui | WRF simulation | 2012/08/07/01 – 2012/08/08/18 | d01, d02, d03 & Study region | $\boldsymbol{X}_{d01}^{t}$, $\boldsymbol{X}_{d03}^{t}$, $\boldsymbol{X}_{\text{pangu}}^{t}$, $\boldsymbol{X}_{\text{LR-obs}}^{t}$, $\boldsymbol{Y}_{\text{LR-osb}}^{t}$, $\boldsymbol{Y}_{\text{ST-osb}}^{t}$ |
| | | Forecast & Obs | 2012/08/04/01 – 2012/08/08/15 | | |
| 1909 | Lekima | WRF simulation | 2019/08/09/07 – 2019/08/10/16 | | |
| | | Forecast & Obs | 2019/08/07/01 – 2019/08/10/15 | | |
| 1918 | Mitag | WRF simulation | 2019/09/30/13 – 2019/10/01/21 | | |
| | | Forecast & Obs | 2019/09/29/13 – 2019/10/02/15 | | |
| 2004 | Hagupit | WRF simulation | 2020/08/03/07 – 2020/08/04/12 | | |
| | | Forecast & Obs | 2020/08/01/16 – 2020/08/05/14 | | |
| 2212 | Muifa | WRF simulation | 2022/09/13/19 – 2022/09/14/18 | | |
| | | Forecast & Obs | 2022/09/12/11 – 2022/09/14/23 | | |
| **Group B:** Events with Forecast & Observation Only | | | | **Used in Correction Module** | |
| 1810 | Ampil | Forecast & Obs | 2018/07/20/23 – 2018/07/22/10 | Study region | $\boldsymbol{X}_{\text{pangu}}^{t}$, $\boldsymbol{X}_{\text{LR-obs}}^{t}$, $\boldsymbol{Y}_{\text{LR-osb}}^{t}$, $\boldsymbol{Y}_{\text{ST-osb}}^{t}$ |
| 1812 | Jongdari | | 2018/08/01/23 – 2018/08/03/10 | | |
| 1814 | Yagi | | 2018/08/11/16 – 2018/08/13/10 | | |
| 1818 | Rumbia | | 2018/08/15/00 – 2018/08/17/10 | | |
| 2106 | In-fa | | 2021/07/23/11 – 2021/07/26/15 | | |
| 2114 | Chanthu | | 2021/09/12/11 – 2021/09/14/15 | | |

* The spatial coverage: d01: 83.91°E–146.09°E & 12.14°N–45.64°N; d02: 116.76°E–123.98°E & 26.06°N–32.24°N; d03: 117.65°E–122.99°E & 26.97°N–31.33°N; Study region: 117.875°E–122.625°E & 27.25°N–31.00°N.

plays two complementary roles: supplying historical inputs ($X_{\text{LR-obs}}$) to characterize forecast biases, and providing spatially continuous labels ($Y_{\text{LR-obs}}$) for training the correction module.

**Ground station observations**: Hourly, 2-min averaged winds from 2,513 stations across Zhejiang Province (Figure 4f) offer high-fidelity, point-scale labels ($Y_{\text{ST-obs}}^{t} \in \mathbb{R}^{S\times2}$). Unlike gridded analyses, which are spatially smoothed, station records directly anchor forecasts at infrastructure-relevant locations, making them indispensable for operational risk assessment.

**Static terrain features**: Elevation data from the 30-m ASTER DEM (Abrams et al., 2020) and land cover classifications are resampled to the 500-m grid of the d03 domain. As shown in Figure 4b, the mountainous and coastal landforms strongly influence wind acceleration and sheltering. These terrain datasets, encoded as invariant conditioning features ($X_{\text{terrain}}$), enable the downscaling module to generalize terrain-induced wind perturbations across diverse TC events.

### 4.2 Preprocessing pipeline

The preprocessing pipeline harmonizes these heterogeneous sources into structured training samples tailored to ACDF's two modules. Its design serves two main goals: (1) aligning spatiotemporal resolutions with the learning objectives of each module, and (2) enabling rigorous evaluation of generalization on unseen storms. Dataset usage is summarized in Table 4.

**Cross-validation and partitioning**: ACDF is trained and validated under a leave-one-storm-out (LOSO) protocol. In each fold, one of the five Group A WRF-simulated TCs—Haikui (2012), Lekima (2019), Mitag (2019), Hagupit (2020), or Muifa (2022)—is withheld for testing, while the remaining events are split into training (85%) and validation (15%) sets. This design maximizes use of limited events while ensuring evaluation under previously unseen tracks, intensities, and terrain interactions. The Hagupit case study (Section 7) corresponds to the fold in which Hagupit was withheld, providing a strict out-of-sample test.

**Downscaling module pre-training**: The downscaling task is posed as a spatial super-resolution problem, requiring paired coarse–fine wind fields that explicitly resolve terrain effects. For each timestep $t$ in the Group A storms, overlapping patches are extracted from d01 (12.5 km) and paired with d03 (500 m) outputs. Each coarse input patch ($X_{\text{d01}}^{i,t} \in \mathbb{R}^{5\times5\times2}$) is matched with its corresponding fine-resolution label ($X_{\text{d03}}^{i,t} \in \mathbb{R}^{101\times101\times2}$) and concatenated with a terrain patch ($X^{i}_{\mathrm{terrain}} \in \mathbb{R}^{101\times 101\times 2}$). Using a patch size of 0.5° × 0.5° and a stride of 0.25°, this procedure yields 35,028 coarse–fine sample pairs. The large sample count arises from patch extraction across both space and time, ensuring dense, physically consistent supervision for terrain-aware refinement.

**Table 4.** Preprocessed dataset feeds into ACDF's two modules.

| Usage | Dataset | Type | Spatiotemporal resolution | Tensor |
|---|---|---|---|---|
| Correction module – Training input | Pangu-Weather forecast | AI-based global forecast | 1 h, 12.5 km | $\boldsymbol{X}^{t}_{\mathrm{pangu}} \in \mathbb{R}^{(H+\tau)\times N\times 2}$ |
| | HRCLDAS analysis (history period) | Gridded observation | 1 h, 12.5 km | $\boldsymbol{X}^{t}_{\mathrm{LR-obs}} \in \mathbb{R}^{H\times N\times 2}$ |
| Correction module – Training label | HRCLDAS analysis (forecast period) | Gridded observation | 1 h, 12.5 km | $\boldsymbol{Y}^{t}_{\mathrm{LR-obs}} \in \mathbb{R}^{\tau\times N\times 2}$ |
| | Weather stations | In situ observations | 1 h, point scale | $\boldsymbol{Y}^{t}_{\mathrm{ST-obs}} \in \mathbb{R}^{\tau\times S\times 2}$ |
| Downscaling module – Pretraining input | WRF-d01 simulation | NWP simulation | 1 h, 12.5 km | $\boldsymbol{X}^{i,t}_{\mathrm{d01}} \in \mathbb{R}^{5\times 5\times 2}$ |
| | ASTER DEM & land cover | Static geospatial data | 500 m | $\boldsymbol{X}^{i}_{\mathrm{terrain}} \in \mathbb{R}^{101\times 101\times 2}$ |
| Downscaling module – Pretraining label | WRF-d03 simulation | NWP simulation | 1 h, 500 m | $\boldsymbol{Y}^{i,t}_{\mathrm{d03}} \in \mathbb{R}^{101\times 101\times 2}$ |

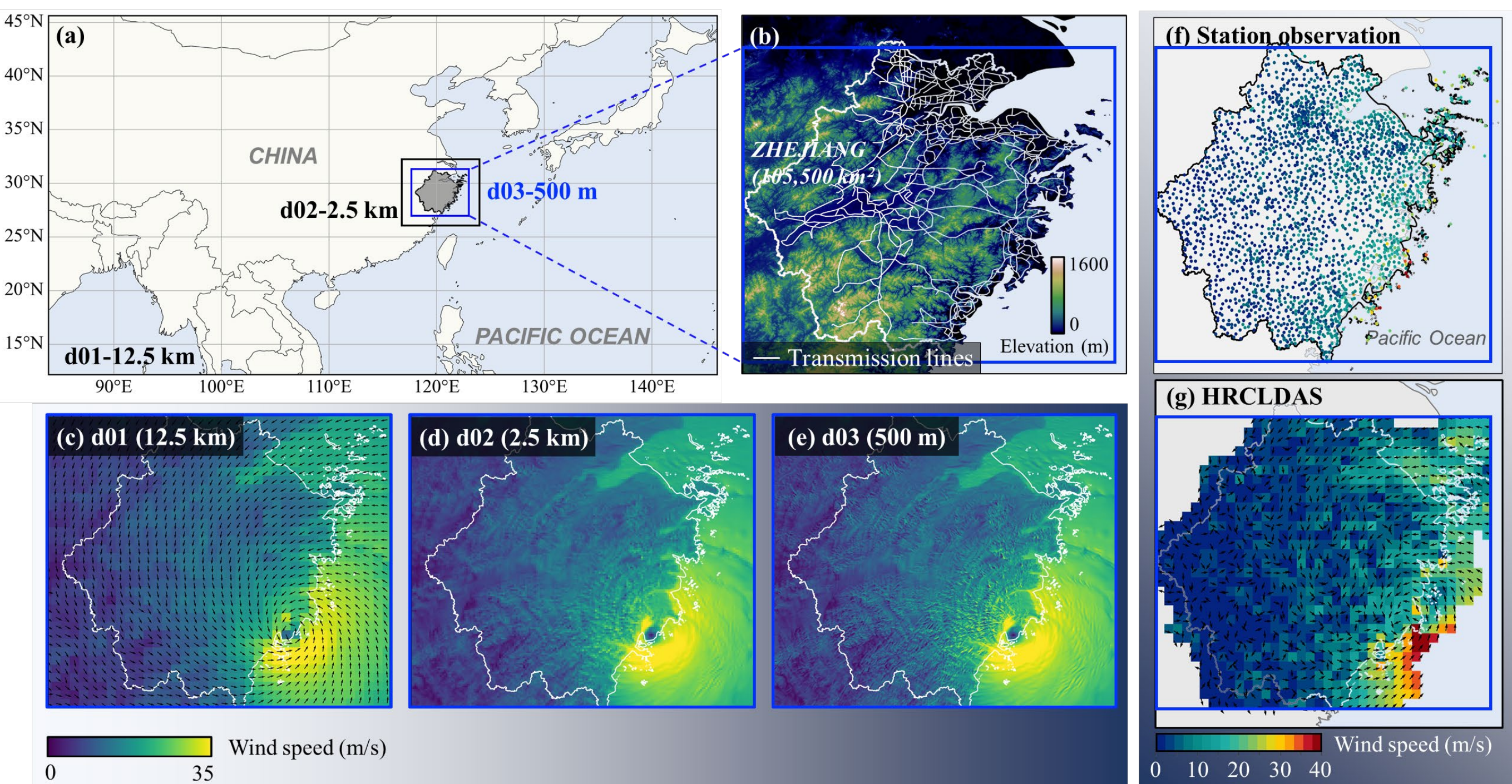

**Fig. 4.** Overview of the study domain and datasets: (a) nested WRF simulation domains (d01: 12.5 km, d02: 2.5 km, d03: 500 m) centered on Zhejiang Province; (b) digital elevation model (DEM) overlaid with the provincial transmission grid; (c-e) d01, d02 and d03 wind field simulations for 2019/08/09/18 UTC; (f) spatial distribution of 2,513 ground weather stations; and (g) Computational domain constrained by station coverage using HRCLDAS wind fields (illustrated with winds at 2019/08/09/18 UTC as an example).

**Correction module end-to-end training**: The correction task is formulated as a spatiotemporal sequence-to-sequence problem. Using all 11 TCs, sliding windows of $(H+\tau)$ hours are constructed, consisting of $H$ hours of historical data and a $\tau$-hour forecast horizon. Inputs include Pangu-Weather forecast ($X_{\mathrm{pangu}} \in \mathbb{R}^{(H+\tau)\times N\times 2}$) for the full $(H+\tau)$ -hour sequence and HRCLDAS gridded analyses ($X_{\mathrm{LR-obs}} \in \mathbb{R}^{H\times N\times 2}$) for the $H$-hour historical segment. Supervised labels are provided by HRCLDAS fields ($Y_{\mathrm{LR-obs}} \in \mathbb{R}^{\tau\times N\times 2}$) and point-scale station observations ($Y_{\mathrm{ST-obs}} \in \mathbb{R}^{\tau\times S\times 2}$) over the $\tau$-hour forecast period. This setup produces 570 spatiotemporal samples, as summarized in Table 4.

Together, these preprocessing steps yield two complementary training sets: (i) dense, terrain-resolved coarse–fine patches for the downscaling module, and (ii) temporally coherent forecast–observation sequences for the correction module. By aligning heterogeneous datasets with the learning objectives of each stage, the pipeline operationalizes ACDF's decoupled design and establishes a structured foundation for model training. Section 5 builds directly on this foundation to describe implementation and optimization.

# 5 Model Architecture and Development

Building on the formulation in Section 3 and the datasets in Section 4, this section details how ACDF is operationalized into a working two-stage system. Rather than treating forecast correction and terrain-aware refinement as a single task, ACDF separates them into complementary modules: one that assimilates observations to reduce storm-scale biases, and another that reconstructs fine-scale, terrain-driven variability. This design not only sharpens model interpretability but also ensures forecasts that are both dynamically consistent and operationally actionable. By modularizing these functions, each stage can be independently validated—storm-scale corrections against coarse analyses and terrain refinements against high-resolution labels.

## 5.1 Terrain-aware downscaling module

The downscaling module functions as a spatial super-resolution network, designed to transform a coarse-resolution (12.5 km) wind field into a physically plausible, high-resolution (500 m) forecast that explicitly accounts for topographical effects ($\boldsymbol{X}_{\text{terrain}}$). This step is critical because infrastructure exposure is highly sensitive to localized accelerations and sheltering effects that are invisible at coarse resolution.

The module is built on a hierarchical encoder-decoder architecture, structurally similar to a U-Net, but it substitutes standard convolutional layers with more powerful Locally-enhanced Window (LeWin) Transformer blocks (Liu et al., 2021; Wang et al., 2022). As detailed in Figures 5c and 5f, this architecture partitions the feature map into non-overlapping windows and computes self-attention exclusively within these local regions, drastically improving computational efficiency while effectively modeling complex spatial dependencies. To better capture fine-grained local patterns, a depth-wise convolutional block is integrated within the feed-forward network (FFN) of each transformer block.

The workflow proceeds as follows (Figure 5c):

- **Input projection** – The coarse wind field $\boldsymbol{X}_{\text{d01}}^{i,t} \in \mathbb{R}^{5\times5\times2}$ is bilinearly interpolated to match the high-resolution terrain grid (500 m, $\mathbb{R}^{101\times101\times2}$) and concatenated with static terrain features $\boldsymbol{X}_{\text{terrain}}^{i} \in \mathbb{R}^{101\times101\times2}$. Both inputs are projected into a shared high-dimensional space.
- **Encoder** – Stacks of DS blocks progressively reduce spatial resolution while increasing channel depth, extracting multi-scale representations.
- **Decoder** – Symmetrical upsampling stages restore spatial resolution, with skip connections from the encoder to preserve high-frequency spatial details critical for terrain effects.

  **Output projection** – A final deconvolution and fully connected layer project the features back to a 2-channel $(U, V)$ wind field $\widehat{\boldsymbol{Y}}_{\text{d03}}^{i,t} \in \mathbb{R}^{101\times101\times2}$.

This hierarchical design ensures that storm-scale circulation patterns are preserved while local terrain effects are superimposed, maintaining both physical plausibility and geographic realism. The module is pre-trained using a mean squared error (MSE) loss against the WRF-simulated reference field:

$$\mathcal{L}_{ds} = \frac{1}{\tau n H' W'} \sum_{i=1}^{n} \sum_{t=1}^{\tau} \left\| \widehat{\boldsymbol{Y}}_{\text{ds}}^{i,t} - \boldsymbol{Y}_{\text{d03}}^{i,t} \right\|^2 \tag{4}$$

where $\tau$ is the forecast horizon, $H'$ and $W'$are the dimensions of the high-resolution grid, and $i$ is the patch index of the total $n$ patches.

## 5.2 Spatiotemporal correction module

The correction module addresses the bias-correction problem described in Section 3.3, targeting systematic forecast errors in coarse-scale AIWP wind fields. It operates on an $(H+\tau)$-hour window comprising $H$ hours of historical observations and forecasts, followed by $\tau$ hours of future forecasts. Specifically, the past forecast–observation pairs are used to learn recent forecast discrepancies, while the future forecasts provide the trajectory to be corrected, as shown in Figure 5a. This design allows recent observational history to directly anchor bias corrections, ensuring the corrected forecasts remain dynamically consistent with the evolving storm environment.

The temporal–spatial relationships are modeled as follows:

1. **Spatiotemporal Graph Attention Network (STGAT)** – Models spatially heterogeneous and temporally evolving forecast errors. Using the $H$-hour historical segment, the STGAT preliminary estimates error patterns and extrapolates them to the forecast horizon, producing a bias field for the full $(H+\tau)$-hour sequence.
2. **Temporal Convolution (TC) block** – Uses gated dilated inception layers (Figure 5d) to capture temporal patterns. The inception structure applies multiple 1D convolutional filters of different kernel sizes in parallel, enabling the model to discover temporal patterns over various time scales simultaneously. Dilation exponentially increases the receptive field with network depth, allowing the model to efficiently process long input sequences and capture long-term dependencies without a sharp increase in computational cost.

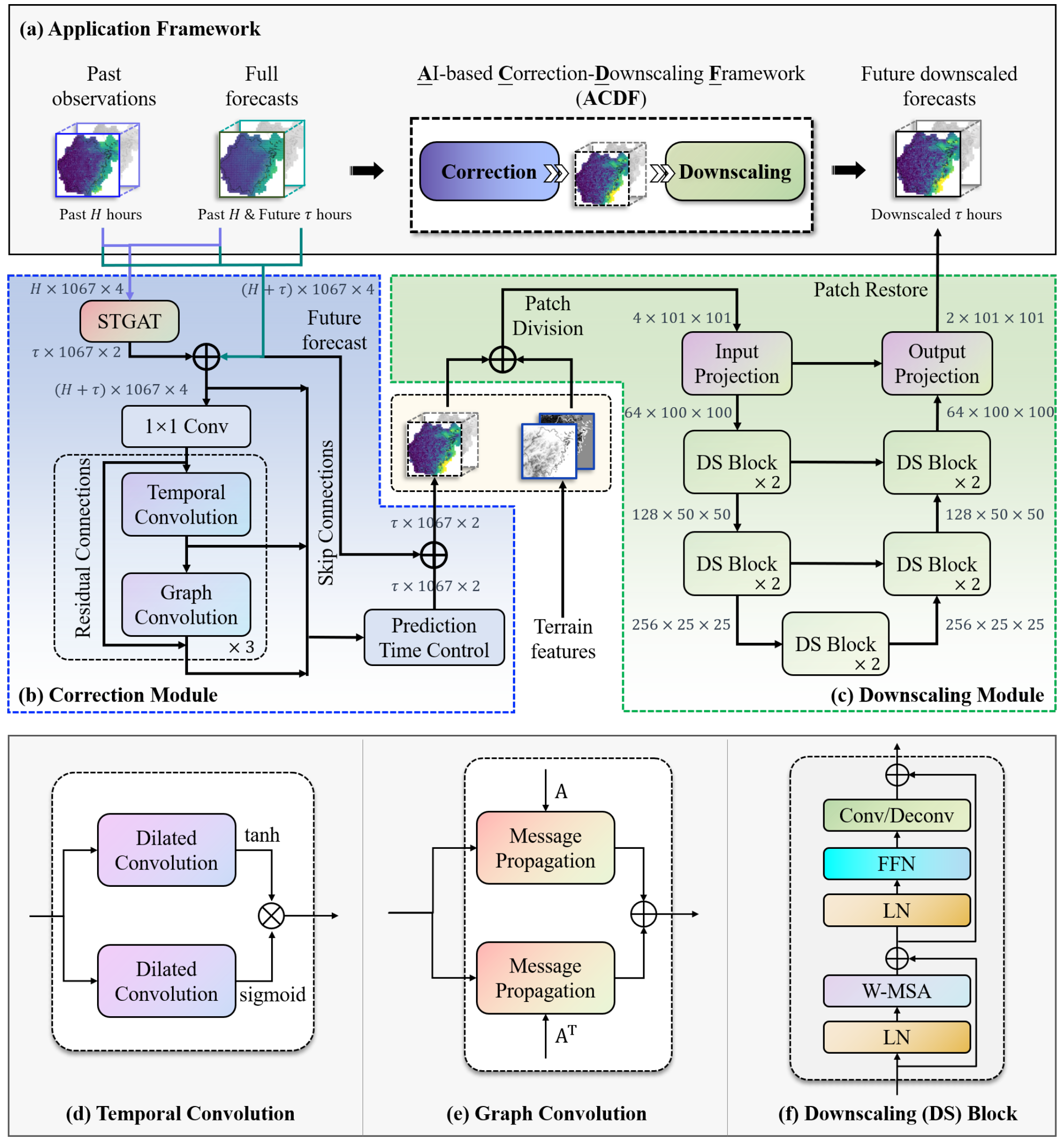

**Fig. 5.** (a) An overview of the ACDF application framework; The neural network architecture implemented in ACDF consists of (b) correction module and (c) downscaling module. (d) Temporal convolution block and (e) graph convolution block of the correction module. (f) The downscaling (DS) block in the downscaling module, consisting of normalization layer (LN), non-overlapping Window-based Multi-head Self-Attention (W-MSA), Feed-Forward Network (FFN) and convolution or deconvolution (Conv/Deconv) layer.

3. **Graph Convolution (GC) block** – Propagates information across the grid according to the learned adjacency matrix (Figure 5e). It employs a mix-hop propagation mechanism, which fuses information from a node's neighborhood to model spatial dependencies (Wu et al., 2020). This design effectively handles the flow of information between grid points while mitigating the over-smoothing problem common in deep GNNs, thereby preserving important local variations.

These TC and GC blocks are interleaved and stacked, producing a $\tau$-hour correction sequence $\Delta\boldsymbol{X}_{corr}$. This correction is applied via a residual connection to generate the bias-corrected coarse field: $\widehat{\boldsymbol{X}}_{\text{corr}} = \boldsymbol{X}_{\text{pangu}} + \Delta\boldsymbol{X}_{corr}$. Together, these components allow the model to capture error patterns that evolve across both time and space — a necessity for reducing intensity biases in AIWP forecasts.

### 5.3 Joint end-to-end optimization

Training follows a sequential, dependency-aware strategy:

1. Pre-train $\mathcal{F}_2$ (downscaling module) using the WRF dataset and Equation 4. Once stable, its parameters $\theta_2$ are frozen.
2. Train $\mathcal{F}_1$ (correction module) with the frozen downscaling module appended, ensuring bias correction is optimized for its impact on the final high-resolution forecast.

The joint optimization is governed by a dual-objective loss:

$$\mathcal{L}_{total} = \mathcal{L}_{grid} + \alpha\mathcal{L}_{station} \tag{5}$$

$$\mathcal{L}_{grid} = \frac{1}{\tau N}\sum_{t=1}^{\tau}\left\|\widehat{\boldsymbol{X}}_{\text{corr}}^{t} - \boldsymbol{Y}_{\text{LR-obs}}^{t}\right\|^{2} \quad (6)$$

$$\mathcal{L}_{station} = \frac{1}{\tau S}\sum_{t=1}^{\tau}\left\|\widehat{\boldsymbol{Y}}_{\text{ds}}^{t} - \boldsymbol{Y}_{\text{ST-obs}}^{t}\right\|^{2} \quad (7)$$

where $\mathcal{L}_{grid}$ enforces meteorological fidelity at the coarse grid scale, and $\mathcal{L}_{station}$ enforces point-level accuracy at weather stations—directly relevant for infrastructure risk. The balance parameter α was selected using validation-set sensitivity analysis and set to 0.01 (Section 6.4; Appendix B). In combination, these two objectives balance large-scale meteorological fidelity with local, infrastructure-relevant accuracy, ensuring the final forecasts are both scientifically credible and operationally useful.

### 5.4 Implementation details

For the correction task, the historical input window was set to $H = 6$ hours and the forecast horizon to $\tau = 12$ hours, balancing timely correction with operational usability. The correction module uses three interleaved TC and GC layers with a dropout rate of 0.3.

The correction and downscaling modules contain 3.82M and 3.73M trainable parameters, respectively, keeping the overall framework lightweight. The framework was implemented in PyTorch and trained on a NVIDIA GeForce RTX 4090 GPU. Training proceeded in two phases: 1) Downscaling pre-training: 200 epochs, Adam optimizer, batch size 32, learning rate $2 \times 10^{-4}$ with a step decay; 2) Correction training: 400 epochs, Adam optimizer, batch size 12, learning rate $2 \times 10^{-3}$, early stopping with patience of 20 epochs.

For the ablation in Section 6.4, the soft-freeze and full-fine-tuning variants share all correction-training settings and differ only in $\mathcal{F}_2$: soft freeze updates only its output head (learning rate $1 \times 10^{-5}$), full fine-tuning updates all of $\mathcal{F}_2$ ($1 \times 10^{-6}$). This staged strategy directly operationalizes the decoupled ACDF design introduced in Section 3, translating architectural choices into a training workflow that aligns with both physical reasoning and practical deployment.

## 6 Model Performance Evaluation

The effectiveness of ACDF is evaluated through rigorous out-of-sample tests designed to stress its ability to generalize across diverse TC conditions. Independent test cases are drawn from five severe events withheld under the leave-one-storm-out (LOSO) protocol, ensuring that performance is always measured on storms unseen during training.

The evaluation spans five dimensions central to infrastructure-scale wind forecasting: (i) event-level accuracy and generalizability, assessed through LOSO testing; (ii) temporal stability, testing whether forecast quality is preserved over a 12-hour horizon; (iii) terrain sensitivity, quantifying the added value of the downscaling module in complex landscapes; (iv) robustness under station-only assimilation, evaluating model performance when HRCLDAS input is replaced by interpolated station data; (v) and computational efficiency, reflecting feasibility of real-time operations.

A central component of the analysis is an ablation test, comparing the full ACDF with a correction-only variant (ACDF-Corr.) to isolate the contribution of the terrain-aware downscaling module. We also include ACDF-S, which assimilates only station-interpolated data instead of HRCLDAS, to test robustness when dense gridded analyses are unavailable. Comparisons with raw Pangu-Weather forecasts and HRCLDAS analyses further benchmark ACDF against both operational forecasts and observation-based references. Station observations are used as the primary independent reference, while HRCLDAS is used to assess consistency with gridded mesoscale analyses.

For consistency across baselines, two standard metrics are employed:

- Mean Error (ME): Quantifies systematic bias (positive = overestimation; negative = underestimation for wind speed).
- Mean Absolute Error (MAE): Captures the magnitude of deviations from station observations and serves as the principal measure of accuracy.

Where relevant, mean ± standard deviation values for MAE and ME are reported, calculated across stations and multiple forecast initialization times.

### 6.1 Aggregate event-level accuracy and generalizability

Table 5 reports station-scale prediction errors for five independent TC events under the LOSO protocol. ACDF attains the best overall accuracy, with an average MAE of 1.374 m/s, improving upon the baseline Pangu forecast (2.245 m/s) by 38.8%. It also achieves station-scale errors comparable to the mesoscale reference HRCLDAS (1.404 m/s on average), indicating that the two-stage correction–downscaling design effectively closes the gap between coarse global guidance (Pangu) and observation-assimilated products (HRCLDAS) while operating at far lower computational cost. Event-wise reductions versus Pangu are consistent—46.6% (Muifa), 41.1% (Lekima), 40.2% (Haikui), 35.1% (Mitag), and 31.2% (Hagupit)—demonstrating robustness across tracks and intensities.

Ablation results clarify each module's contribution. The correction-only variant (ACDF-Corr.) lowers average MAE to 1.447 m/s (–35.5% vs Pangu), showing that assimilating recent observations efficiently removes storm-scale biases in intensity and structure. Adding the terrain-aware module (i.e., ACDF) yields the lowest errors (1.374 m/s), capturing ridge speed-up and valley sheltering that are invisible at coarse resolution.

**Table 5.** Station-scale mean ± standard deviation of MAE and ME for five LOSO test storms (Best mean values in **bold**; second-best in underline (for ME, closeness to zero is considered "better"))

| Testing Event (ID) | Model | $MAE_{spd}$ (m/s) | $ME_{spd}$ (m/s) | $MAE_{dir}$ (°) |
|---|---|---|---|---|
| Haikui (1211) | Pangu | 2.374±1.890 | 1.821±2.427 | 38.668±40.559 |
| | ACDF-Corr. | 1.600±1.696 | **-0.073±2.330** | 37.341±40.528 |
| | ACDF-S | 1.570±1.653 | -0.307±2.259 | 38.305±41.684 |
| | ACDF (Ours) | **1.419±1.506** | -0.168±2.062 | **36.008±39.857** |
| | HRCLDAS (Target) | 1.705±1.783 | 0.190±2.459 | 39.282±42.441 |
| Lekima (1909) | Pangu | 2.705±2.052 | 1.948±2.780 | 42.266±43.585 |
| | ACDF-Corr. | 1.631±1.813 | -0.505±2.386 | 39.906±42.535 |
| | ACDF-S | **1.590±1.725** | -0.638±2.258 | 41.606±43.352 |
| | ACDF (Ours) | 1.594±1.760 | **-0.385±2.343** | **39.529±42.434** |
| | HRCLDAS (Target) | 1.600±1.811 | -0.230±2.406 | 39.154±42.074 |
| Mitag (1918) | Pangu | 2.001±1.697 | 1.315±2.270 | 47.194±44.504 |
| | ACDF-Corr. | 1.329±1.501 | 0.582±1.919 | 43.747±43.661 |
| | ACDF-S | **1.290±1.437** | -0.633±1.825 | 48.074±45.680 |
| | ACDF (Ours) | 1.299±1.452 | **-0.488±1.886** | **43.403±43.484** |
| | HRCLDAS (Target) | 1.252±1.502 | -0.189±1.946 | 40.321±41.913 |
| Hagupit (2004) | Pangu | 1.792±1.604 | 1.128±2.124 | 43.866±42.581 |
| | ACDF-Corr. | 1.296±1.418 | -0.718±1.782 | 42.257±42.528 |
| | ACDF-S | 1.273±1.374 | -0.696±1.739 | 49.249±46.981 |
| | ACDF (Ours) | **1.233±1.343** | **-0.585±1.727** | **40.810±41.949** |
| | HRCLDAS (Target) | 1.166±1.328 | -0.429±1.715 | 38.822±40.656 |
| Muifa (2212) | Pangu | 2.452±1.597 | 1.827±2.287 | 45.873±43.613 |
| | ACDF-Corr. | 1.360±1.510 | -0.604±1.940 | 42.450±43.103 |
| | ACDF-S | 1.341±1.447 | -0.713±1.839 | 44.753±44.457 |
| | ACDF (Ours) | **1.309±1.431** | **-0.456±1.885** | **41.448±42.480** |
| | HRCLDAS (Target) | 1.287±1.456 | -0.394±1.903 | 40.325±42.812 |
| Average | Pangu | 2.245±1.820 | 1.584±2.418 | 43.314±43.087 |
| | ACDF-Corr. | 1.447±1.604 | -0.501±2.102 | 40.994±42.483 |
| | ACDF-S | 1.416±1.544 | -0.597±2.008 | 44.354±44.684 |
| | ACDF (Ours) | **1.374±1.515** | **-0.421±2.002** | **40.088±42.080** |
| | HRCLDAS (Target) | 1.404±1.605 | -0.212±2.121 | 39.476±41.899 |

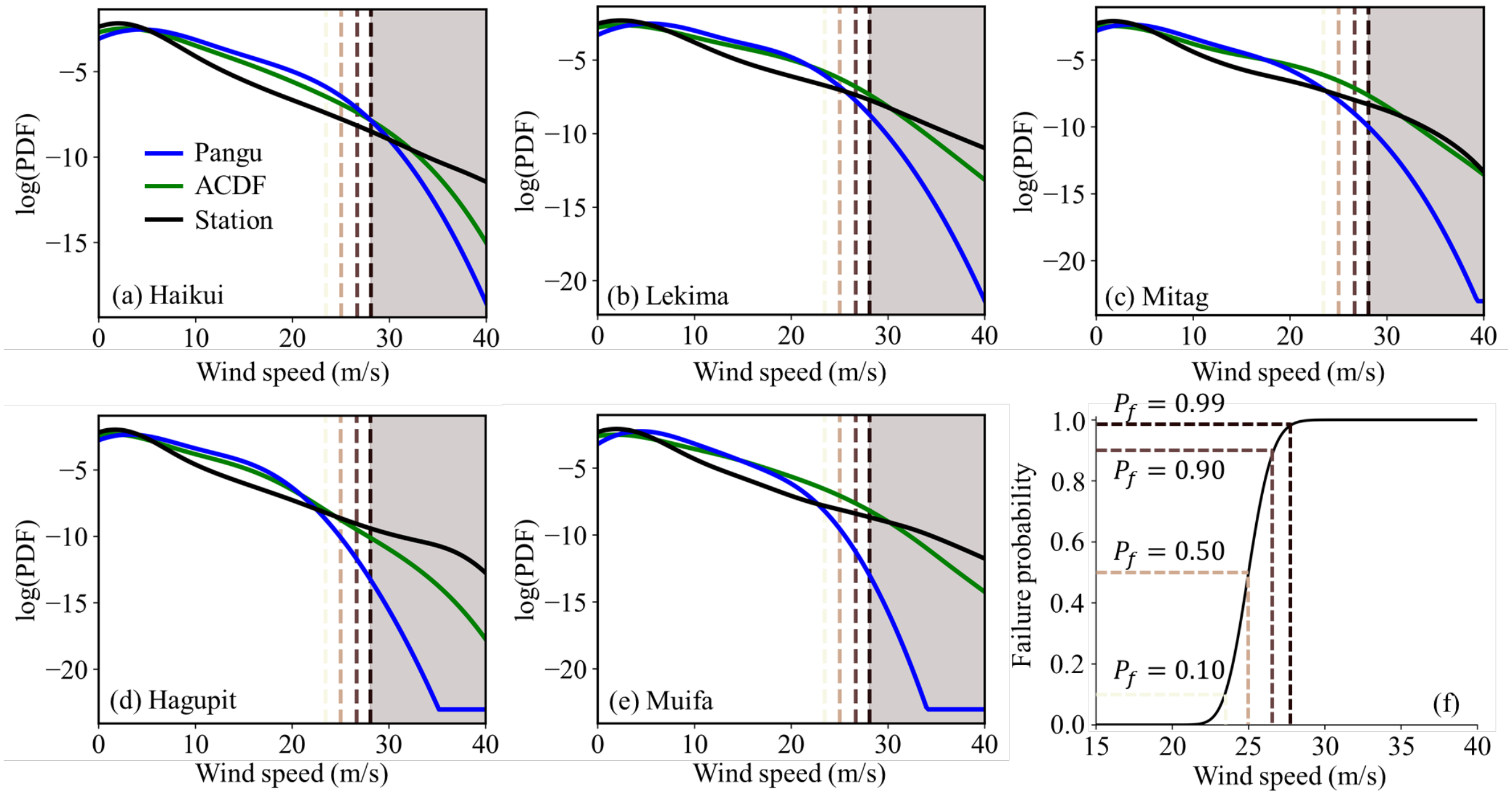

**Fig. 6.** (a-e) Probability density functions for all station samples across five severe TC events, showing that ACDF aligns closer with observations and reduces the high-wind underestimation seen in Pangu forecasts; and (f) a typical fragility function for the transmission towers.

Biases are likewise mitigated. The average ME decreases from 1.584 m/s (Pangu) to –0.421 m/s (ACDF), indicating a more balanced representation across weak- and strong-wind regimes. Wind-direction errors also improve modestly at the aggregate level (from 43.314° to 40.088°, with HRCLDAS at 39.476°), which is operationally adequate given the known uncertainties in mesoscale directional labels.

Regarding dispersion, the reported standard deviations (STDs) are large across all methods, reflecting strong spatiotemporal heterogeneity in storm impacts and label variability; importantly, ACDF's MAE dispersion (1.515 m/s) is lower than Pangu's (1.820 m/s) and slightly better than HRCLDAS (1.605 m/s). Two factors explain this behavior without undermining the central finding: (i) ACDF's gains over Pangu are substantial and systematic, especially in the high-wind regime that drives fragility; and (ii) the mesoscale target (HRCLDAS) itself exhibits sizable STDs, so ACDF's near-parity with HRCLDAS suggests that much of the remaining dispersion is label-side rather than model-induced. Overall, ACDF delivers lower mean error, reduced bias, and tighter spread than the raw AI forecast, providing stable and asset-relevant guidance for downstream risk assessment.

Figure 6 further illustrates the distributional benefits of ACDF. In the high-wind regime (>20 m/s)—critical for transmission tower vulnerability—Pangu consistently underestimates observed winds, while ACDF significantly narrows this gap. The alignment with observed distributions, particularly in the upper tail where fragility functions steeply increase failure probability, demonstrates ACDF's operational value. Taking a representative fragility curve as shown in Figure 6f, the probability of tower failure rises steeply once wind speeds exceed 20 m/s, reaching 0.5 at around 25 m/s. When these fragility thresholds are overlaid on the probability distributions of each TC (Figures 6(a–e)), it is clear that Pangu rarely captures the density of extreme winds, while ACDF produces distributions that closely match observations.

The advantage of ACDF becomes even more evident in the gray-shaded region corresponding to total tower failure (100% probability). Here, Pangu's forecasts rarely reach the critical thresholds, while ACDF succeeds in reproducing the observed density of extreme winds. This distinction is operationally crucial: because tower failure probability is accumulated over time (details in Section 6.3), repeated underestimation of high winds—even by small margins—can lead to severe misjudgments in system-level risk. By more faithfully capturing these upper-tail extremes, ACDF ensures that forecasts translate into more reliable and realistic failure probability assessments.

Taken together, these results demonstrate that ACDF not only corrects TC-scale biases and lowers mean error but also captures the high-wind extremes most relevant for infrastructure risk, offering clear advantages over the Pangu baseline.

### 6.2 Temporal correction stability

Beyond event-level accuracy, the temporal stability of forecasts is equally critical for operational reliability. Forecasts that fluctuate sharply over time can undermine user confidence and complicate decision-making, even if their average accuracy appears acceptable. To evaluate this aspect, we focus on an 18-hour analysis window spanning from 6 hours before landfall to 12 hours after, which captures the most hazardous stage of the event when wind speeds typically exceed 15 m/s and forecast errors have the greatest impact. Within this window, multiple forecast cycles are considered, each providing a 12-hour forecast horizon. This setup allows us to assess not only pointwise performance but also the consistency of forecast quality over successive updates as the storm evolves.

As shown in Figure 7a, both Pangu and ACDF maintain relatively steady MAE values across the 12-hour forecast horizon. However, their absolute accuracy diverges. Pangu's errors remain consistently above 3 m/s, whereas ACDF sustains values below 2 m/s throughout. Pangu also shows an anomalous trend of decreasing error with lead time. This trend aligns with post-landfall weakening in the storms' intensity time series and reflects tropical cyclone evolution rather than genuine forecast skill: early timesteps coincide with peak intensity when biases are greatest, while later timesteps fall into the decay phase when weaker winds naturally reduce error magnitudes. A regime-stratified check further confirms that this decrease in high-wind MAE does not correspond to a systematic increase in low-wind error; low-wind MAE remains approximately stable across lead times for both Pangu-Weather and ACDF (Appendix C).

The differences become more pronounced when focusing on high-wind conditions (Figure 7b). Here, Pangu's performance is not only less accurate but also highly unstable, with pronounced oscillations in mean error and a much broader spread across forecast cycles. In contrast, ACDF delivers a substantially narrower error distribution and smoother temporal evolution, consistently maintaining mean MAE values below 4 m/s. The interquartile ranges (IQRs) plotted in Figure 7 further support this finding: ACDF exhibits tighter IQRs than Pangu across all lead times, indicating reduced variability and greater robustness from one forecast cycle to the next.

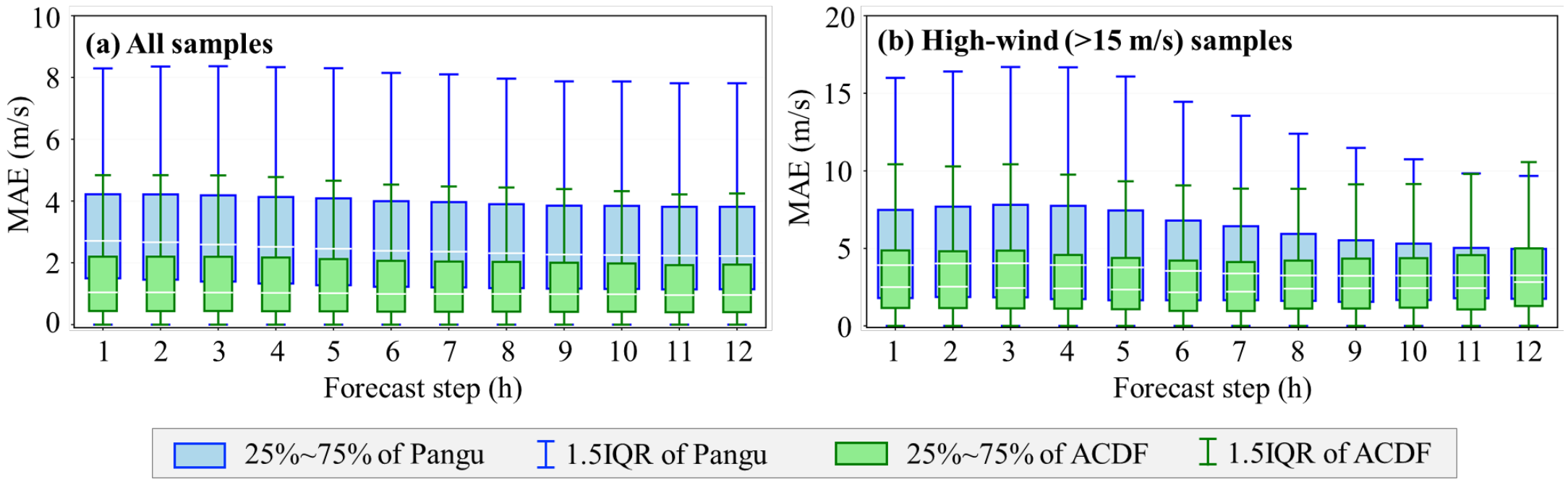

**Fig. 7.** Temporal evolution of wind speed MAE for (a) all samples and (b) high-wind samples over a 12-hour forecast horizon. The shaded region denotes the interval of one standard deviation (STD) about the mean (Mean ± STD) for the sample data.

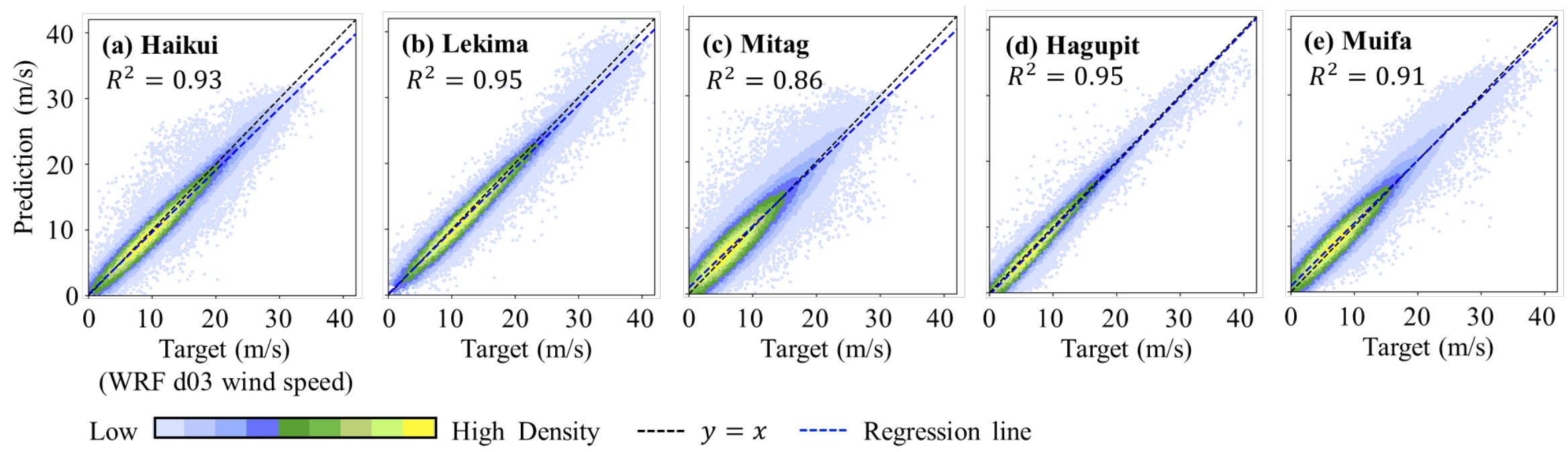

**Fig. 8.** LOSO tests of downscaling predicted versus WRF wind speeds for the (a) Haikui, (b) Lekima, (c) Mitag, (d) Hagupit, and (e) Muifa events.

### 6.3 Terrain sensitivity

The terrain-stratified analysis uses the same 18-hour window as in Section 6.2 —from 6 h before landfall to 12 h after—so that temporal and spatial variations are examined under identical storm stages. Figures 8(a-e) presents the regression results between pre-trained downscaling model predictions and WRF-simulated 500 m reference winds for the five independent TC cases. Across Haikui, Lekima, Mitag, Hagupit, and Muifa, the coefficient of determination ($R^2$) reaches 0.93, 0.95, 0.86, 0.95, and 0.91, respectively. These consistently high correlations demonstrate that the terrain-aware downscaling module reconstructs fine-scale wind structures with remarkable fidelity and generalizes well across storms of differing tracks and intensities. The strong cross-event agreement confirms that ACDF does not simply memorize terrain patterns from training storms but robustly captures the physical relationship between large-scale flow and local topographic modulation.

Using the Topographic Position Index (TPI) (Weiss, 2001), sites are categorized as valleys (< −50 m), flats (−50 m $\leqslant TPI \leq$ 50 m), and ridges (> 50 m) to quantify terrain-specific performance (Figure 9). Pangu's errors show little differentiation among terrain classes, whereas ACDF progressively refines predictions: bias correction (ACDF-Corr.) reduces MAE by 46%, 36%, and 16% for valleys, flats, and ridges, and adding the downscaling module further cuts errors by 59%, 47%, and 32%, respectively. Although relative gains are smaller on ridges, they are especially valuable because these zones host the highest winds and most vulnerable assets. Moreover, ACDF markedly narrows the IQR of MAE across all classes, indicating reduced dispersion and greater forecast reliability. Altogether, the framework achieves terrain-consistent, temporally stable, and cross-event-generalizable performance—transforming coarse AI forecasts into physically credible, infrastructure-relevant wind fields.

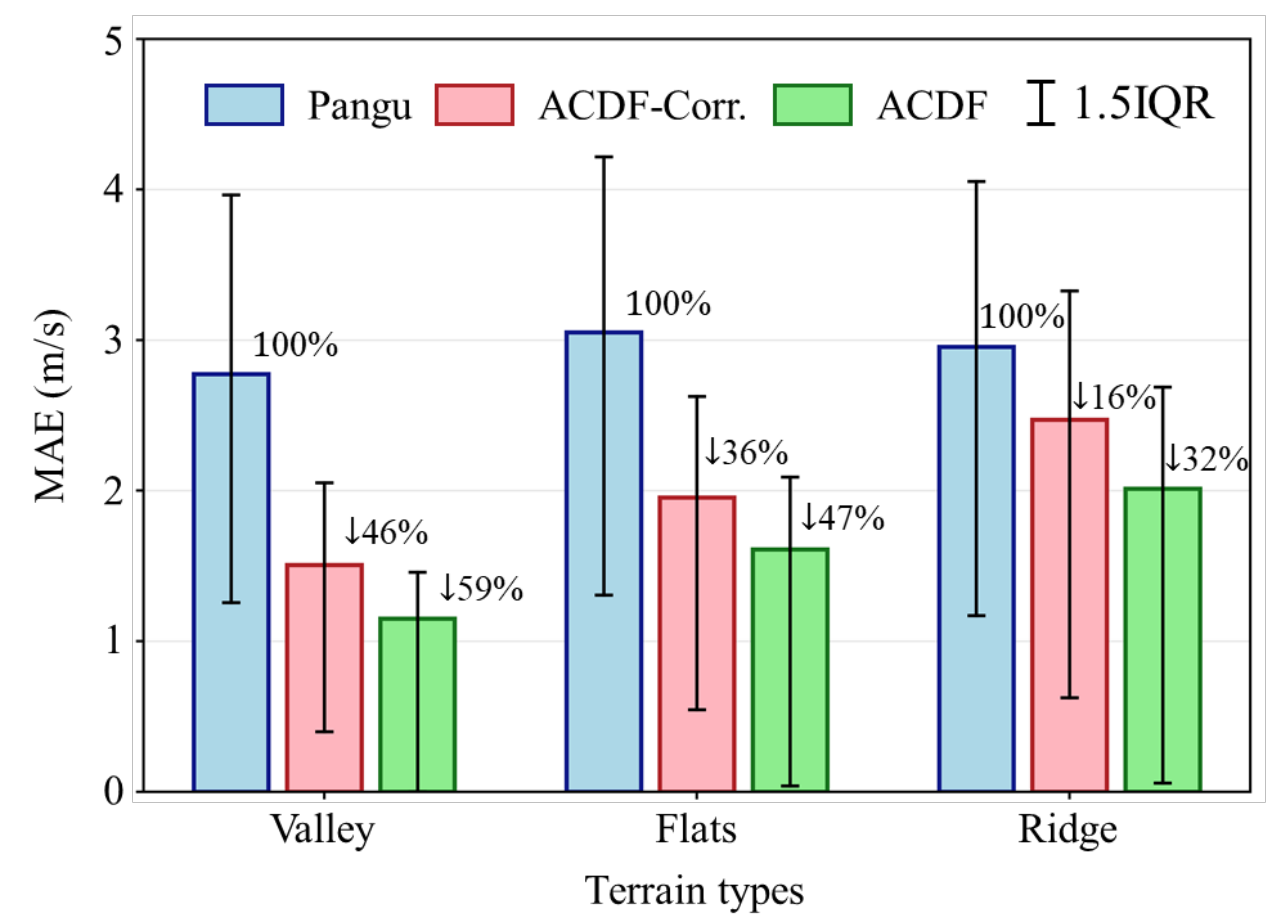

**Fig. 9.** Wind-speed MAE stratified by terrain types: valleys ($TPI$ < –50 m), flats (–50 m ≤ $TPI$ ≤ 50 m), and ridges ($TPI$ > 50 m).

### 6.4 Validation of the Sequential Correction–Downscaling Strategy

The downscaling module $\mathcal{F}_2$ is pre-trained using paired WRF-d01–WRF-d03 simulations, but during inference it is driven by bias-corrected Pangu-Weather fields. This transfer setting raises three linked questions: whether $\mathcal{F}_2$ should remain frozen during Stage 2, whether the corrected Pangu-Weather fields are compatible with the WRF-d01 input domain used to train $\mathcal{F}_2$, and whether $\mathcal{F}_2$ remains useful when applied directly to corrected Pangu-Weather fields under deployment conditions.

We first compare three Stage-2 training strategies: hard freezing $\mathcal{F}_2$, soft freezing $\mathcal{F}_2$, and low-learning-rate full fine-tuning of $\mathcal{F}_2$ (Figure 10(a)). Hard freezing yields the lowest and most stable station-scale wind-speed errors, with the advantage most pronounced in the high-wind regime (>15 m/s), which is most relevant to fragility-based transmission risk. This behavior is expected because Stage 2 provides coarse HRCLDAS fields and sparse station observations, but no dense 500 m supervision. Relaxing $\mathcal{F}_2$ during Stage 2 therefore risks over-adapting the terrain-response operator to sparse station residuals and weakening the WRF-learned mapping from coarse flow and terrain to fine-scale wind perturbations. Hard freezing also removes $\mathcal{F}_2$ from backpropagation during Stage 2, reducing training cost and improving optimization stability.

The adopted hard-freeze configuration trains stably across all five LOSO folds (Figure 10(b)). Training and validation losses decrease together and plateau without persistent divergence, and early stopping halts each fold near its validation minimum. This indicates that the correction module, with 3.82M trainable parameters, does not show evident overfitting under the available spatiotemporal training sequences.

Under the hard-freeze configuration, the station-weighting coefficient $\alpha$ controls the strength of station-level anchoring relative to gridded-field consistency. Because $L_{grid}$ is computed over a dense spatial field whereas $L_{station}$ is computed only at sparse station locations, $\alpha$ should not be interpreted as the physical importance of station observations. Instead, it balances local point-scale supervision against regional spatial coherence. We therefore selected $\alpha$ using validation-set sensitivity analysis within each LOSO training fold, rather than using the withheld test storm. As summarized in Appendix B, $\alpha = 0.01$ yields the lowest and most stable validation error among the candidate values $\{0, 0.01, 0.05, 0.1, 0.3\}$. Setting $\alpha = 0$ slightly underuses station observations, whereas larger values degrade validation accuracy monotonically and destabilize convergence, as sparse station residuals increasingly dominate the composite objective.

Hard freezing is viable only if the corrected fields driving $\mathcal{F}_2$ at inference remain compatible with the WRF-d01 fields used to train it. We test this directly in Figure 11 by comparing WRF-d01, raw Pangu-Weather, HRCLDAS, and the corrected Pangu-Weather field produced by the correction module. Correction shifts the wind-speed distribution from raw Pangu-Weather toward WRF-d01, reducing the Wasserstein-1 distance from 3.61 to 1.49 m/s. It also reduces the kinetic-spectral distance from 1.84 to 0.93, indicating improved agreement in the spatial-scale energy structure of the coarse wind field. The spectra are especially consistent at low wavenumbers, where storm-scale structure is represented; the remaining small-scale discrepancy corresponds to the terrain-conditioned variability that $\mathcal{F}_2$ is designed to reconstruct.

We further evaluate $\mathcal{F}_2$ under direct deployment inputs (Table 6). Applying the frozen $\mathcal{F}_2$ directly to raw Pangu-Weather offers only marginal benefit at the all-station level, reducing MAE from 3.209 to 3.088 m/s (3.8%), and in fact slightly worsens the high-wind regime, from 6.106 to 6.400 m/s. This confirms that terrain-aware downscaling alone cannot remove—and may even amplify—storm-scale bias in the upstream AIWP field, particularly in the high-wind tail

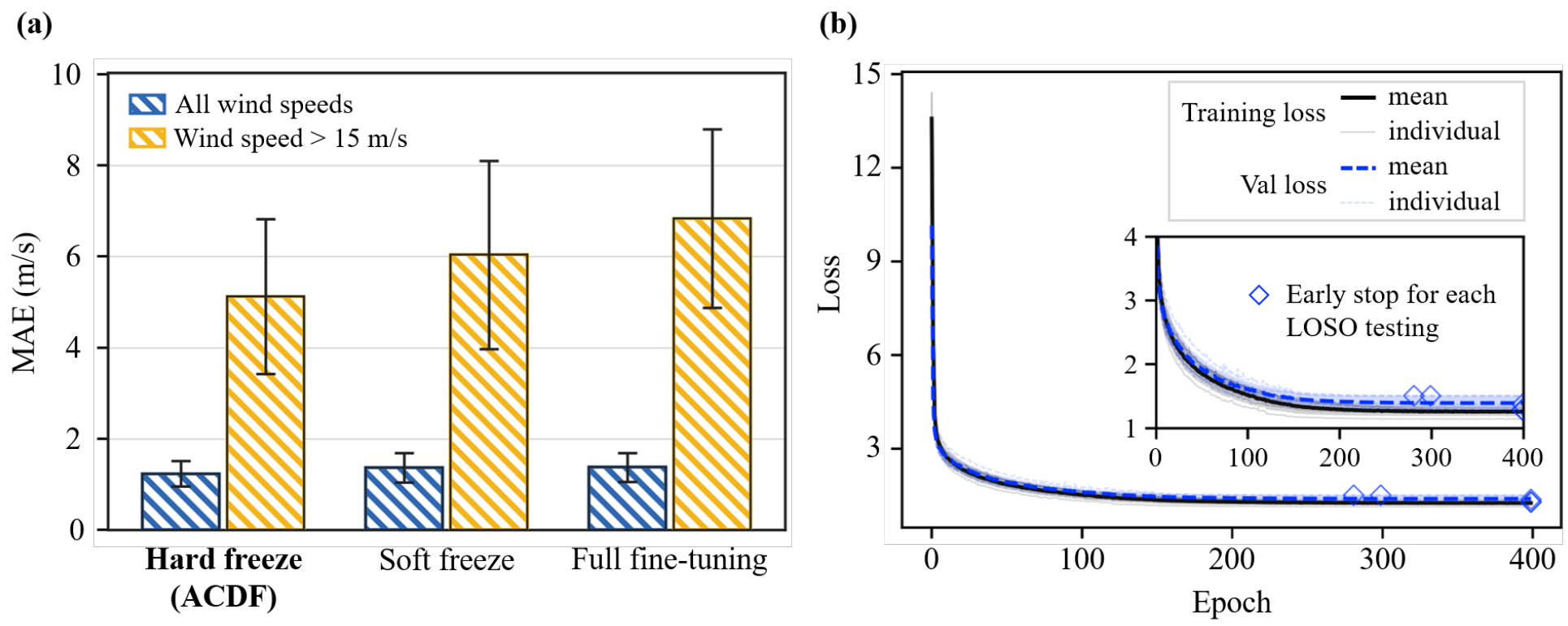

**Fig. 10.** Validation of the sequential training strategy. (a) Comparison of three Stage-2 freezing strategies for the downscaling module—hard freeze (ACDF), soft freeze, and full fine-tuning—in terms of station-scale wind-speed MAE, reported for all wind speeds and for the high-wind regime (>15 m/s). (b) Training and validation loss curves across the five LOSO folds under the adopted hard-freeze configuration.

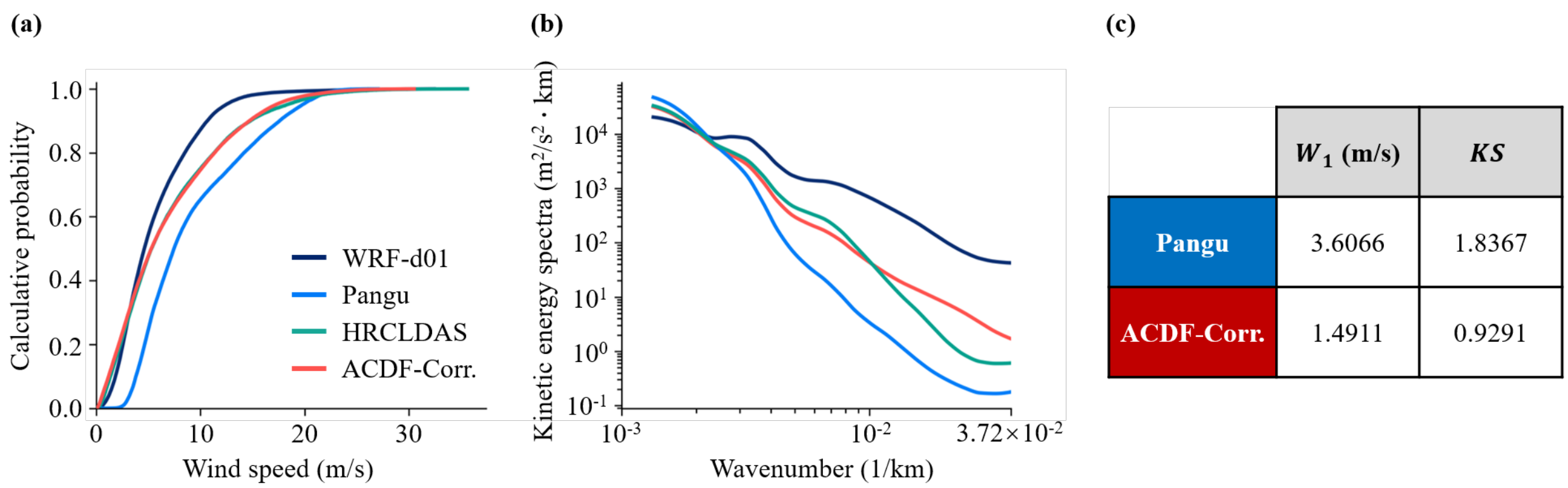

**Fig. 11.** Distributional consistency between $\mathcal{F}_2$'s training input (WRF-d01) and its operational input after correction. (a) Cumulative distribution of wind speed for WRF-d01, Pangu, HRCLDAS, and ACDF-Corr.; (b) corresponding kinetic energy spectra; (c) Wasserstein-1 ($W_1$) and Kinetic Spectra (KS) distances of Pangu and ACDF-Corr. relative to WRF-d01.

**Table 6.** Direct deployment-input evaluation of the frozen downscaling module $\mathcal{F}_2$.

| Variant | Input to $\mathcal{F}_2$ | $\mathcal{F}_2$ used? | All-station MAE (m/s) | High-wind MAE (m/s) | MAE reduction vs. raw Pangu | Interpretation |
|---|---|---|---|---|---|---|
| Raw Pangu | — | No | 3.209 | 6.106 | — | Raw AIWP baseline |
| Raw Pangu + $\mathcal{F}_2$ | Raw Pangu | Yes | 3.088 | 6.400 | 3.8% | Downscaling raw inputs barely helps and worsens the high-wind tail |
| ACDF-Corr. | — | No | 2.024 | 5.982 | 36.9% | Correction substantially removes storm-scale bias |
| ACDF | Corrected Pangu | Yes | 1.919 | 5.684 | 40.2% | Proposed deployable configuration; $\mathcal{F}_2$ adds terrain detail after correction |

**Notes:** Values are averaged over the five LOSO withheld Group-A storms. High-wind MAE is computed for station-time samples with observed wind speed greater than 15 m/s.

that governs fragility-based risk. Correction, by contrast, substantially reduces this bias: the correction-only variant lowers all-station MAE to 2.024 m/s (36.9%). Driving the same frozen $\mathcal{F}_2$ with the corrected field (full ACDF) then gives the best performance, reducing all-station MAE to 1.919 m/s (40.2%) and, crucially, improving the high-wind MAE from 5.982 to 5.684 m/s. In other words, $\mathcal{F}_2$ contributes additional skill in the critical high-wind regime only when it operates on a corrected input, confirming that correction moves Pangu-Weather into a regime where the WRF-trained terrain-response operator remains effective.

Together, these results validate the sequential correction–downscaling strategy. The correction module first moves raw Pangu-Weather toward the WRF-d01 input domain in both wind-speed distribution and spectral structure. Hard freezing then preserves the WRF-learned terrain-response mapping in $\mathcal{F}_2$, while direct deployment-input testing confirms that $\mathcal{F}_2$ contributes additional skill only after the upstream AIWP field has been corrected. This supports the central design choice of ACDF: correcting storm-scale bias before applying terrain-aware refinement.

### 6.5 Robustness under station-only data assimilation (ACDF-S)

Having established ACDF's skill with multi-source assimilation, we next stress-test its robustness in data-sparse conditions. To test robustness in data-sparse settings, we replace HRCLDAS inputs with station-interpolated fields while keeping the architecture unchanged. As presented in Table 5, across LOSO events, ACDF-S achieves an average MAE of 1.416 m/s—a 36.9% reduction vs Pangu and within 0.042 m/s of the full ACDF (1.374 m/s). Bias control is similar (ME −0.597 m/s for ACDF-S vs −0.421 m/s for ACDF). Directional errors are larger for ACDF-S (44.354°), reflecting the loss of dense gridded constraints, but wind-speed skill—the primary driver of fragility—remains strong. The small performance gap confirms practical implementation capacity across heterogeneous data regimes.

In practice, ACDF can operate in two modes: a baseline, station-only mode (ACDF-S) for sparse regions and a multi-source mode (ACDF) when gridded analyses are available. Thus, operational agencies can tailor assimilation strategies to available resources—leveraging simple setups where data are sparse, and maximizing accuracy when dense products exist.

## 6.6 Computational efficiency

A defining advantage of the ACDF is its computational efficiency, which makes real-time operational deployment feasible. For a single forecast cycle covering a 12-hour lead time, the complete correction and downscaling process requires only approximately 25 seconds of inference time on an NVIDIA 4090 GPU, with the downscaling module accounting for 95% of this cost. By contrast, generating a comparable high-resolution forecast using the traditional physics-based WRF model would require several hours of runtime. This three-orders-of-magnitude speedup transforms what was once a computationally prohibitive task into one suitable for rapid decision-making during evolving TC events.

These efficiency gains do not come at the expense of accuracy. Across evaluations, ACDF consistently delivers forecasts that are accurate, unbiased, and reliable, while also sustaining skill across diverse terrains and storm conditions. The ability to produce terrain-aware, infrastructure-scale wind fields in seconds rather than hours underscores ACDF's practical value: it bridges the gap between research-grade models and real-time operational needs, establishing itself as a field-ready framework for infrastructure-scale wind forecasting under TC conditions.

# 7 Case Study: From Pangu-Weather Forecasts to Sub-Kilometer Wind And Risk Forecasting of Zhejiang's Power Transmission System During Typhoon Hagupit

## 7.1 Event setup and forecast configuration

To illustrate ACDF's end-to-end operational capacity, we present Severe Typhoon Hagupit (ID: 2004) in Zhejiang Province as a representative use case. The case traces the full workflow—assimilation and bias correction of AIWP mesoscale forecasts, sub-kilometer terrain-aware downscaling, and dynamic tower- and line-level risk estimation—showing how ACDF converts raw AIWP output into infrastructure-specific intelligence fast enough for operations during a rapidly evolving landfalling TC.

As shown in Figure 12a, Pangu-Weather was run in three forecast cycles on 3 August 2020 at 00:00, 06:00, and 12:00 UTC. At each cycle, ACDF (i) ingests the past 6 h and the next 12 h of raw Pangu mesoscale forecasts, (ii) assimilates 6 h of HRCLDAS analyses together with 2,513 ground-station observations, (iii) performs bias correction at 12.5 km, and (iv) downscales to 500-m resolution. This tightly coupled process produces future 12-hour terrain-resolved, unbiased wind field forecasts in real time and drives risk forecasts for 780 transmission lines, spanning about 9,500 km across ~105,500 km² of Zhejiang.

The lead hours (shown in Figures 12(a,b)) for the four key analysis times used in Section 7.2 are:

- 12:00 UTC (pre-landfall): 12 h lead from Cycle I and a 6 h lead from Cycle II.
- 18:00 UTC (close to landfall): 12 h lead from Cycle II and a 6 h lead from Cycle III.
- 19:00 UTC (landfall): 7 h lead from Cycle III.
- 22:00 UTC (inland decay): 10 h lead from Cycle III.

These staggered lead times let us assess how ACDF behaves both earlier (12 h) and closer (6–10 h) to impact as new observations arrive—an operationally critical property during rapid TC evolution.

For each analysis time (12:00, 18:00, 19:00, 22:00 UTC), Figure 12b juxtaposes two forecasts when available—one issued in an earlier cycle (longer lead) and one in a later cycle (shorter lead)—against HRCLDAS. This layout isolates the effect of lead time while holding the valid time fixed.

**12:00 UTC** (pre-landfall; Cycle I 12 h vs. Cycle II 6 h). ACDF-Corr. reduces Pangu's storm-scale bias; full ACDF then projects those corrections into a terrain-aware 500 m field, sharpening coastal gradients and ridge–valley contrasts without spurious extremes. The shorter-lead Cycle II forecast best matches HRCLDAS.

**18:00 UTC** (close to landfall; Cycle II 12 h vs. Cycle III 6 h). The lead–skill contrast strengthens. Pangu underestimates the compact coastal core; ACDF-Corr. improves intensity and placement; ACDF at 6 h lead (Cycle III) aligns most closely with HRCLDAS in both magnitude and footprint.

**19:00 UTC** (landfall; Cycle III 7 h). With fresh assimilation, ACDF maintains the observed amplitude over the right-hand semicircle of the TC track and captures localized coastal maxima missed by Pangu.

**22:00 UTC** (inland decay; Cycle III 10 h). ACDF preserves the >10 m/s footprint and terrain-conditioned peaks near 20 m/s, showing the model maintains forecast accuracy across the event lifecycle.

## 7.2 Wind field enhancement

Across timestamps, shorter-lead cycles consistently yield higher skill, especially in the high-wind tail, while longer-lead forecasts are smoother and left-biased in intensity. This demonstrates the operational value of real-time cycling: each 6-hour update tightens TC-scale bias through assimilation, and ACDF then projects those corrections into sub-kilometer, terrain-aware structure.

Figure 13 steps from province to asset scale to show why bias correction and downscaling are both required for credible line-failure forecasting. In the province scale column

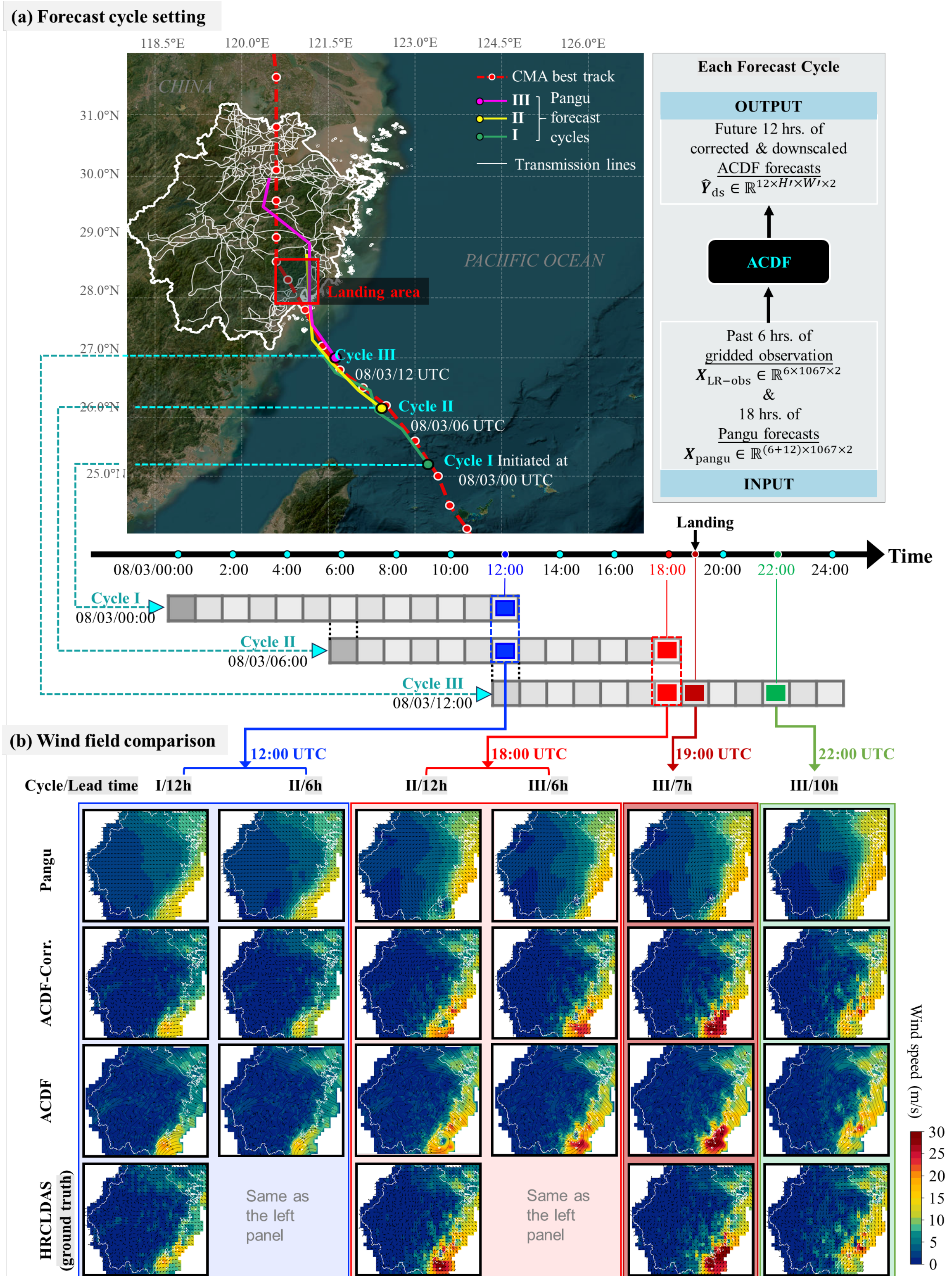

**Fig. 12.** (a) CMA best track and Pangu forecast tracks for Typhoon Hagupit at three forecast cycles; (b) comparison of Pangu, ACDF-Corr., ACDF and HRCLDAS wind fields at four key times with different lead times.

(Figures 13(a-c)), the raw Pangu field is smooth and weak, blurring the compact coastal wind core at landfall. ACDF-Corr. corrects storm-scale intensity and placement, and the full ACDF preserves those corrections while adding terrain-conditioned structure at 500-m resolution. In the landfall zone column (Figures 13(d-f)), station dots expose the differences. Pangu concentrates values in the 0–10 m/s range and misses the coastal high-wind corridor. ACDF-Corr. restores mesoscale gradients consistent with the evolving storm, and ACDF resolves ridge speed-up and valley sheltering along the mountainous coastline—features required to explain observed extremes. The payoff appears at the line segments column (Figures 13(g-i)). Pangu and ACDF-Corr. both yield nearly uniform loading along spans; only ACDF produces spatial variability at the tower–span scale—localized maxima on exposed ridges and minima in sheltered valleys—that drives fragility-based failure probabilities. The topography/line overlay (Figure 13k) clarifies the orographic controls, while the tower-level distributions (Figure 13j) confirm that Pangu deviates noticeably from station observation; ACDF matches stations and recovers the >10 m/s tail that governs failure risk.

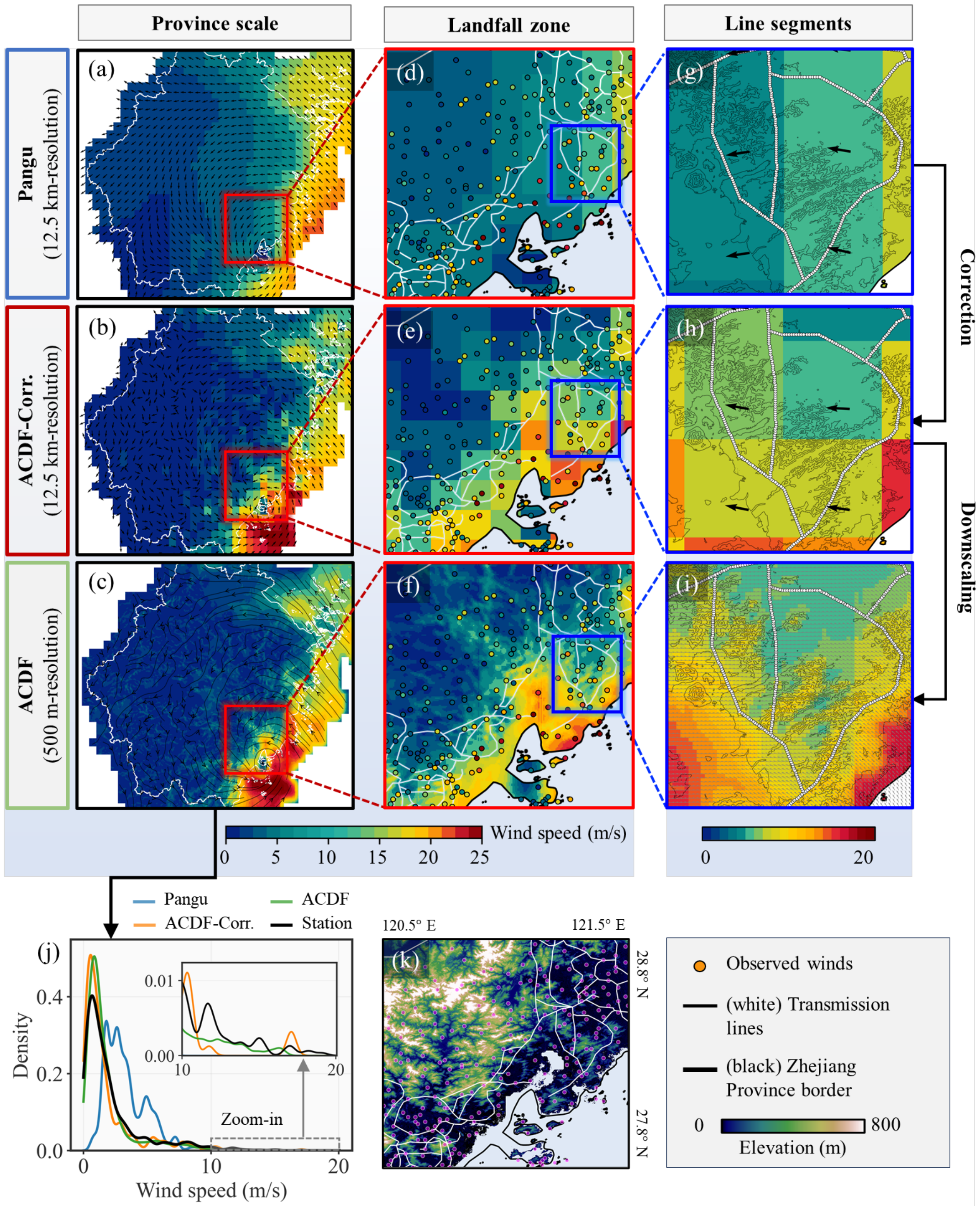

**Fig. 13.** Multi-scale comparison at 18:00 UTC, 3 Aug 2020 (close to landfall). Columns: Province scale—(a) Pangu (12.5 km), (b) ACDF-Corr. (12.5 km), (c) ACDF (500 m); Landfall zone—(d) Pangu, (e) ACDF-Corr., (f) ACDF with station observations; Line segments—(g) Pangu, (h) ACDF-Corr., (i) ACDF along a transmission corridor; (j) tower-level wind-speed distributions (Pangu, ACDF-Corr., ACDF, station observations); (k) Topography with transmission lines. ACDF retains storm-scale corrections and adds 500 m terrain-modulated structure (ridge speed-up, valley sheltering), recovers the observed >10 m/s tail, and yields asset-scale variability necessary for credible tower/line failure forecasting.

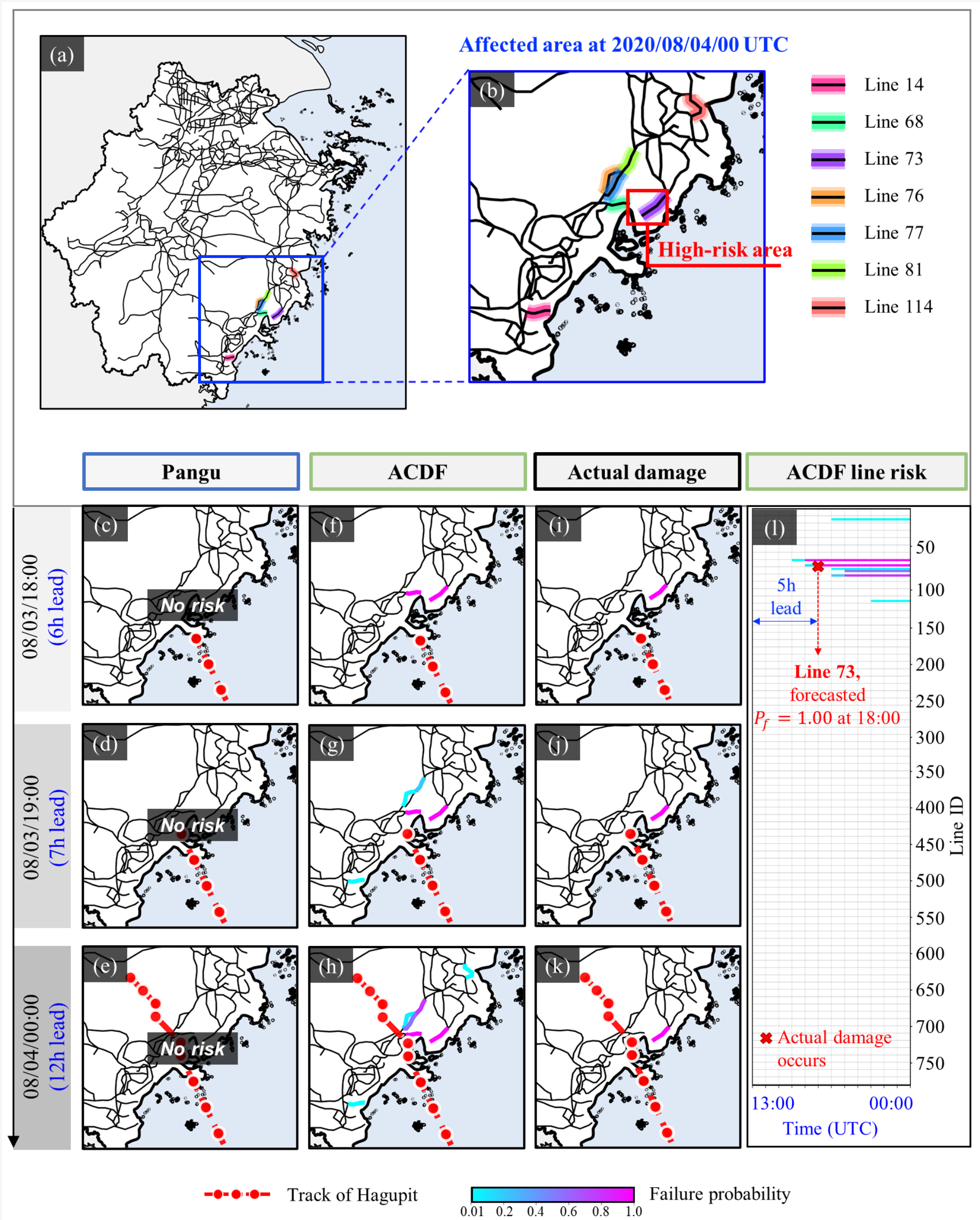

**Fig. 14.** Comparative risk assessment for the Zhejiang transmission system during Typhoon Hagupit. (a) Province-wide map of line-level failure probabilities and (b) zoom near the landfall zone. (c–k) Line failure probabilities at three key valid times—18:00 (close to landfall), 19:00 (landfall), and 00:00 UTC (final forecast time)—comparing (c, d, e) Pangu and (f, g, h) ACDF against (i, j, k) recorded line failures; and (l) Time-varying failure probabilities across all lines.

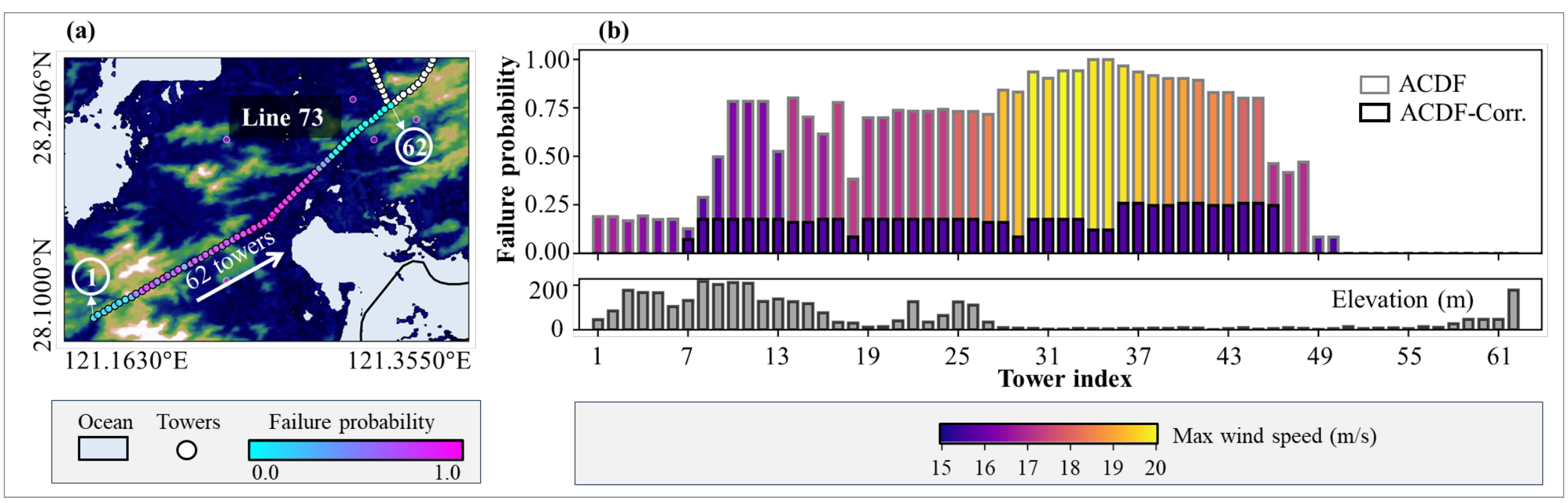

**Fig. 15.** (a) Surrounding topography and final failure probabilities of the Line 73; (b) Final failure probabilities, maximum wind speeds from 13:00 UTC August 3rd to 00:00 UTC August 4th and elevations of towers in Line 73.

Together, these panels demonstrate how the two ACDF modules act on different stages of the forecasting-to-risk pathway. The raw Pangu-Weather field retains storm-scale bias and therefore provides a distorted basis for line-level risk estimation. ACDF-Corr. reduces this large-scale bias, while the full ACDF further restores terrain-modulated 500 m variability needed to differentiate exposure along transmission corridors. This progression from Pangu-Weather to ACDF-Corr. and then to ACDF shows that the framework does not simply sharpen the original coarse forecast; instead, it first corrects upstream forecast errors and then adds local terrain structure, reducing their propagation into asset-level risk estimates.

### 7.3 Infrastructure risk forecasting

The objective is then to translate ACDF's sub-kilometer winds into actionable, real-time risk for the transmission network. We couple the 500-m wind fields to fragility models for Zhejiang's ~9,500 km of 780 transmission lines supported by 32,458 towers, producing tower- and line-level failure probability forecasts with lead times of up to 12 h.

**Risk mapping**. For each tower $j$ at hour $t$ during the 12-hour forecast period, ACDF supplies 10-m wind speed $v_{j,t}$ and attack angle $\theta_{j,t}$ . An angle-dependent lognormal fragility converts $(v_{j,t}, \theta_{j,t})$ to instantaneous tower failure probability $P'_{j,t}$ after marginalizing capacity uncertainty $R$ by Monte Carlo. Temporal cumulative risk uses a survival update—once a tower fails it remains failed—yielding the cumulative probability $P_j$. Line risk treats each line $i$ as a series system, $P_{i,t}^{Line} = 1 - \Pi_{j \in line_i}(1 - P_{j,t})$. Full formulas (as well documented in Huang & Wang, 2024; Xue et al., 2020) and settings follow standard reliability practice are provided in Appendix A (Equations A1–A5; Table A1). For Hagupit we use a single low-capacity tower class and conditional independence; utilities can substitute type-specific fragilities and dependence structures without changing the workflow.

**System-level patterns**. Figure 14 visualizes line-level failure probabilities from Cycle III at 18:00 (close to landfall), 19:00 (landfall), and 00:00 UTC (final forecast time). Because Pangu strongly underestimates winds, no lines exceed the operational threshold ($P_i^{Line} = 0.01$) in Figures 14(c–e), highlighting its unsuitability for network-scale risk prediction. In contrast, ACDF (Figures 14(f-h)) delineates a coherent coastal high-risk corridor aligned with the right-hand semicircle of maximum TC winds and tracks its evolution through the event. Among 780 lines, ACDF elevates only a small subset as potentially critical—precisely the triage operators need. Notably, it flags the actual failed line (ID 73) (Figures 14(i-k); $P_{73}^{Line} = 1.00$) and highlights additional corridors (IDs 68, 77, 81) where conductor galloping/over-stress was reported, while Pangu misses all. These are not spurious "false alarms" but physically plausible risk signals. A mild positive bias is operationally acceptable, as it focuses attention on truly hazardous corridors within a large network. The time series in Figure 14l shows line failure probabilities rising and then stabilizing as high-wind exposure accumulates—behavior consistent with the survival update and observed impacts—and illustrates lead-time utility: successive updates converge quickly on a stable, high-confidence set of critical members rather than "everything high-risk" or "nothing at risk." Notably, for line 73, the recorded actual failure occurred at 18:00 UTC, precisely when the ACDF-predicted failure probability reached 1.00, offering a valuable 5-hour lead time for grid emergency decision-making and demonstrating strong temporal alignment with the actual damage event.

**Asset-level differentiation** (Figure 15). Over complex landfall terrain, ACDF-Corr. (correction only) yields near-uniform span loads along Line 73 and muted failure probabilities. The full ACDF resolves valley channeling and ridge speed-up, producing localized wind maxima >20 m/s and sharp tower-to-tower contrasts in $P_{j,t}$. In Figure 15(a), the surrounding higher terrain forms a small valley-like corridor that channels winds along the line route, leading to locally amplified exposure, while Figure 15(b) co-plots (along the line) final $P_j$, maximum wind over 13:00–00:00 UTC, and elevation—revealing one-to-one correspondence between orographic exposure, wind peaks, and failure hotspots. This granularity pinpoints which spans/towers warrant pre-emptive switching, targeted inspection, or staged restoration.

**Operational cadence**. Each AIWP forecast-to-risk cycle completes in ~25 s on a single RTX 4090 GPU, providing 12-hour lead time for proactive operations. The same pipeline accommodates AIWP ensemble inputs to generate probabilistic risk products (forecast uncertainty), without architectural changes.

This case study shows a complete path from global forecasts to decisions: ACDF corrects and downscales in seconds and outputs tower- and line-level failure probabilities forecasts with 12-hour lead time, localizing risk hot spots before landfall at the resolution of individual assets, creating concrete opportunities for targeted inspections, pre-emptive switching, and staged restoration. The same workflow generalizes across storms, grids, and data regimes, positioning ACDF as an immediately useful tool for resilience operations.

## 8 Conclusion

This work addresses an implementation gap in climate resilience: turning fast, global AI weather prediction into asset-scale, impact-based intelligence that operators can use.

We introduced the AI-based Correction–Downscaling Framework (ACDF), a modular pipeline that first removes storm-scale biases from AI weather forecasts and then restores sub-kilometer, terrain-conditioned wind structure, thereby combining real-time correction, terrain-aware downscaling, and infrastructure risk translation within a single operational workflow. The resulting 500-m, province-scale wind product is not intended as a standalone resolution advance, but as an infrastructure-oriented representation of terrain-modulated wind variability along transmission corridors.

Across 11 historical typhoons, ACDF reduced station-scale wind-speed MAE by 38.8% relative to Pangu-Weather, achieved station-scale accuracy comparable to HRCLDAS, and delivered terrain-aware 500-m wind fields over a province-scale domain while running in ~25 s per 12-h cycle on a single GPU. By better preserving terrain-induced acceleration and sheltering along coastal ridges, valleys, and transmission corridors, the 500 m fields reduce the smoothing of local wind exposure that can otherwise affect fragility-based failure probabilities and high-risk line prioritization. Ablations confirmed complementary roles: correction fixes intensity and placement errors; downscaling supplies the terrain detail that governs structural loading. ACDF maintained temporal stability across the 12-h horizon, generalized across storms and terrain, and remained effective in a station-only mode (ACDF-S), preserving ~95% of headline gains when gridded analyses were unavailable.

The Typhoon Hagupit (2020) case study demonstrated end-to-end operational value. ACDF reproduced observed high-wind tails, revealed a coherent coastal risk corridor, and flagged the transmission line that later failed, enabling tower- and line-level failure probabilities that aligned with observed impacts. Six-hour cycling provided a clear lead-time/skill trade-off: as new observations arrived, storm-scale bias tightened and those corrections were projected into sub-km structure—precisely the regime that drives asset fragility.

Limitations and priorities for future work remain. Our experiments are region-specific; broader trials are needed to assess transferability across coastal settings and network types. The current workflow is driven by deterministic AI forecasts; propagating meteorological uncertainty through to failure probabilities via ensemble drivers is a natural next step. Fragility modeling used a single conservative tower class and assumed conditional independence; adopting type-specific fragilities and spatial/temporal dependence models would sharpen risk estimates. Beyond wind-driven outages, ACDF's modular design is readily extensible to multi-hazard contexts (e.g., rainfall-induced failures) and to other networks (transportation, water, communications).

Despite these caveats, ACDF offers a practical pathway to operational, impact-based early warning. It is fast, interpretable, and scalable; it couples naturally with ensemble AI drivers, outage propagation models, and restoration planning. By bridging global AI forecasts to actionable, asset-level risk, ACDF advances how civil infrastructure prepares for and rides through tropical cyclones—supporting decisions that safeguard lives and reduce economic losses under a changing climate.

## Acknowledgment

This research was co-supported by Zhejiang Department of Science & Technology (Grant No. 2024C03255) and by the State Grid Zhejiang Electric Power Co., Ltd. (Grant No. 5211DS25000L).

# Appendix A: Tower and Line Risk Mapping

## A.1 Tower failure probability

This section formalizes the tower-level risk as a time-evolving probability of collapse driven by forecast wind loading. The formulation distinguishes between (i) the instantaneous, conditional failure probability at a given time step $t$, and (ii) the cumulative failure probability up to that time, accounting for survival in previous steps. The resistance $R$ of a tower—its capacity against wind-induced collapse—is treated as an uncertain but time-invariant random variable over the forecast horizon (e.g., 12 h). Wind load $S_t$ denotes the effective demand at time $t$ obtained by mapping the wind speed (and wind attack angle) simulated by the ACDF to the positions of the transmission towers, and calculate the wind load distributed along the height of the transmission tower based on the wind-receiving area.

**Independent failure probability**. If one is only interested in the failure probability under a representative load level $S$ (e.g., a peak or design wind state), the probability of failure $P_f$ is given by the total probability with respect to resistance uncertainty:

$$\begin{aligned} P_f &= P(R < S) \\ &= \sum_{r \in R} P_{f|r} \cdot P(R = r)\,, P_{f|r} = P(r < S) \end{aligned} \tag{A1}$$

**Cumulative probability up to time $t$.** Over a forecast period $\tau$ with time-invariant resistance $R = r$,

$$\begin{aligned} P_{f,t} &= 1 - \sum_{r \in R} P(r \geq S_1, \ldots, r \geq S_t) \cdot P(R = r) \\ &= \sum_{r \in R} P_{f,t|r} \cdot P(R = r) \end{aligned} \tag{A2}$$

Assuming adjacent steps are statistically independent when the interval is several minutes (consistent with turbulence decorrelation under strong winds), the survival update for the conditional probability is:

$$P_{f,t|r} = P_{f,t-1|r} + \left(1 - P_{f,t-1|r}\right) P'_{f,t|r} \tag{A3}$$

where $P'_{f,t|r}$ is probability of tower collapse at time $t$ only under load $S_t$ conditional on $R = r$, which is the fragility function of the tower conditioned on a prescribed $R = r$. Then the $P_{f,t}$ becomes:

$$P_{f,t} = P_{f,t-1} + \sum_{r \in R} \left(1 - P_{f,t-1|r}\right) \cdot P'_{f,t|r} \cdot P(R = r) \tag{A4}$$

These relations follow from the total probability theorem and the independence approximation at large time intervals (e.g., 10-min).

**Fragility function for towers.** The failure probability of the tower up to time $t$, $P_{f,t}$, can be estimated by using Monte

Carlo simulation (MCS) to sample tower resistance ($r$) from their predetermined distributions as previously discussed, then using fragility functions developed for different $r$ values (i.e., $P'_{f,t|r}$) in each time segment for every sample realization. The detailed method of fragility functions development for transmission towers can be found in existing studies (Fu et al., 2019; Mara and Hong, 2013; Tian et al., 2020; Chen et al., 2026). For illustrative purpose in this study, all towers are assumed to be statistically independent and to share the same low-capacity fragility function, which is defined by a set of log-normal distributions $\Phi(\mu_\theta, \sigma_\theta)$ corresponding to different wind attack angles $\theta$. The fragility function is defined as in Table A1.

**Table A1.** Fragility parameters by wind attack angle used in this study

| Attack angle $\theta$ | 0° | 30° | 45° | 60° | 90° |
|---|---|---|---|---|---|
| $\mu_\theta$ | 2.708 | 2.996 | 3.219 | 3.401 | 3.555 |
| $\sigma_\theta$ | 0.03 | 0.03 | 0.03 | 0.03 | 0.03 |

### A.2 Line-level failure probability

A transmission line $i$ is operationally unavailable if any of its supporting tower $j$ collapses. This motivates a series-system abstraction in which the line fails when at least one tower fails:

$$P_{f,t}^{Line,i} = 1 - \sum_{\vec{r} \in \vec{R}} \prod_{j=1}^{n_i} \left(1 - P_{f,t|r_j}^{Tower,j}\right) \cdot P\left(\vec{R} = \vec{r}\right) \quad \text{(A5)}$$

where tower $j$ is the $j$th one in total $n_i$ supporting towers of line $i$; $P_{f,t|r_j}^{Tower,j}$, obtained by Equation A4, is the failure probability of tower $j$ up to time $t$ conditional on $R_j = r_j$; $\vec{r}$ is the correlated samples for all $n_i$ towers realized from the correlated tower resistances, $\vec{R}$, using MCS method.

### A.3 Algorithmic workflow

The inputs and the end-to-end workflow used to convert ACDF wind forecasts into tower- and line-level failure probabilities suitable for operations are summarized as follows:

- **Wind fields**: 500 m resolution, hourly 10-m winds from ACDF over the forecast horizon.
- **Tower geolocation & orientation**: latitude/longitude for each transmission tower, plus line azimuth to compute wind-attack angle at the structure.
- **Line topology**: tower–line membership (series system abstraction: a line is down if any of its towers collapses).
- **Tower fragility and resistance**: angle-dependent lognormal fragility functions $P'_{f,t|r}$ that map wind load to instantaneous collapse probability, as defined in Table A1 in this study. These were developed with both load and resistance uncertainties considered during model development.

Let $t = 1, \ldots, \tau$ index hourly forecast times.

Step 1. Interpolate the ACDF wind vector field to each tower location and decompose it relative to the line azimuth to obtain the attack angle. From wind speed and attack angle, compute the effective wind load on towers (GB 50009-2012, 2012; ASCE, 2015).

Step 2. Evaluate the instantaneous conditional failure probability at time $t$ using the tower-specific fragility as defined earlier (see Equation A1). Update the cumulative (up-to-time-$t$) tower failure probability using the survival-update relation as defined earlier (see Equation A4).

Step 3. Treat each line as a series system of its supporting towers. Aggregate tower probabilities into a line-level failure probability using the complement-of-survival relation as defined earlier (see Equation A5).

Repeat Steps 1–3 for all $t = 1, \ldots, \tau$. The survival update in Step 2 ensures monotonically non-decreasing tower probabilities over time; the series aggregation yields time-evolving line probabilities.

## Appendix B: Sensitivity of the station-loss weight $\alpha$

Figure B1 presents the validation-set sensitivity analysis used to select the station-loss weight $\alpha$ in Equation 5, with candidate values $\alpha \in \{0, 0.01, 0.05, 0.1, 0.3\}$. All curves are computed on the validation subset within each LOSO training fold, never on the withheld test storm, and report a normalized validation MSE under a common criterion so that different $\alpha$ settings are directly comparable.

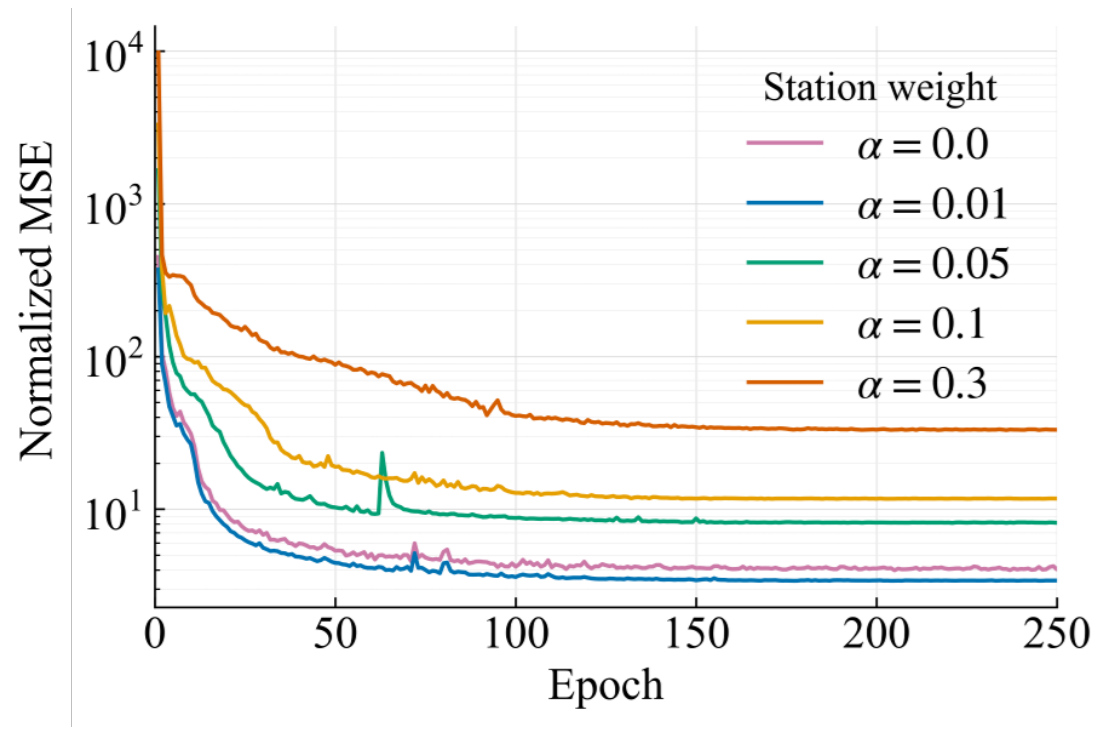

**Fig. B1.** Sensitivity of the station-loss weight $\alpha$ in the correction-module objective.

**Table C1.** Lead-time MAE stratified by observed wind-speed regime.

| Forecast lead time | Pangu low wind <10 m/s | Pangu moderate wind 10–20 m/s | Pangu high wind >20 m/s | ACDF low wind <10 m/s | ACDF moderate wind 10–20 m/s | ACDF high wind >20 m/s |
|---|---|---|---|---|---|---|
| 1 h | 3.03 | 4.09 | 10.52 | 1.60 | 3.67 | 8.38 |
| 2 h | 2.97 | 4.18 | 10.86 | 1.60 | 3.93 | 8.52 |
| 3 h | 2.91 | 4.20 | 10.83 | 1.61 | 4.11 | 8.31 |
| 4 h | 2.86 | 4.19 | 10.73 | 1.62 | 4.23 | 7.97 |
| 5 h | 2.82 | 4.17 | 10.43 | 1.63 | 4.28 | 7.54 |
| 6 h | 2.79 | 4.06 | 9.98 | 1.63 | 4.14 | 7.41 |
| 7 h | 2.77 | 3.94 | 9.47 | 1.63 | 3.89 | 7.82 |
| 8 h | 2.76 | 3.86 | 8.94 | 1.59 | 3.63 | 8.27 |
| 9 h | 2.75 | 3.76 | 8.45 | 1.56 | 3.42 | 8.66 |
| 10 h | 2.75 | 3.71 | 8.02 | 1.52 | 3.26 | 8.80 |
| 11 h | 2.75 | 3.66 | 7.66 | 1.50 | 3.13 | 8.77 |
| 12 h | 2.76 | 3.66 | 7.49 | 1.47 | 3.07 | 8.72 |
| Mean | 2.83 | 3.96 | 9.45 | 1.58 | 3.73 | 8.26 |

**Notes:** Wind-speed regimes are defined using observed station wind speed. Values are averaged over LOSO test storms and station-time samples. The decrease in high-wind MAE with lead time mainly reflects storm weakening and changing sample composition after landfall, while low-wind MAE also decreases lightly rather than increasing.

As shown in Figure B1, $\alpha = 0.01$ converges to the lowest validation error, marginally below the station-free baseline ($\alpha = 0$), while larger values degrade accuracy monotonically—by roughly a factor of two at $\alpha = 0.05$ and an order of magnitude at $\alpha = 0.3$—with slower, less stable convergence. Because $\mathcal{L}_{station}$ is evaluated only at sparse station locations whereas $\mathcal{L}_{grid}$ spans a dense field, an overly large $\alpha$ allows point residuals to dominate the gradient and destabilize the regional wind field. We therefore adopt $\alpha = 0.01$, noting that this small coefficient is a regularization choice balancing local anchoring against spatial coherence, not a measure of the importance of station observations.

## Appendix C: Lead-time MAE stratified by wind-speed regime

Table C1 reports lead-time MAE separately for low-, moderate-, and high-wind station samples. Wind-speed regimes are defined using observed station wind speed. The results show that the decrease in high-wind MAE with lead time does not reflect a compensating increase in low-wind error. Low-wind MAE remains approximately stable across the 12-hour forecast horizon for both Pangu-Weather and ACDF, while the high-wind MAE decreases mainly because the analyzed post-landfall period contains progressively fewer extreme-wind samples as the storm weakens inland.